\documentclass[12pt]{article}
\usepackage{amssymb}
\newcommand{\vek}[1]{\mbox{\bf #1}}
\begin{document}
\hspace*{\fill} LMU--TPW--97/10\\
\hspace*{\fill} hep-th/9704193 \\[3ex]

\begin{center}
\Large\bf
Implications of a Dilaton in Gauge Theory and Cosmology
\end{center}
\vspace{2ex}
\normalsize \rm
\begin{center}
{\bf Rainer Dick}\\[0.5ex] {\small\it
Sektion Physik der Universit\"at M\"unchen\\
Theresienstr.\ 37, 80333 M\"unchen,
Germany}
\vspace{3ex}
\end{center}
\small{\bf Contents}\\[2ex]
\begin{tabular}{rlr}
1. & Introduction & 2  \\
2. & Instantons in gauge theory& 11\\ 
3. & The Kaluza--Klein paradigm & 16 \\   
4. & The dilaton in three dimensions & 24  \\ 
5. & Generalized Coulomb potentials in gauge theory with a dilaton& 27  \\ 
6. & The axidilaton and stabilization of the dilaton 
     in four dimensions & 32 \\ 
7. & The supersymmetric theory & 40\\ 
8. & The dilaton as a dark matter candidate & 42  \\ 
9. & Conclusions and outlook & 51 \\ 
    & Appendix A: Conventions and notation & 52  \\
    & Appendix B: A remark on perturbative aspects
        of the axidilaton &\qquad 54 \\
    & References & 56  \\
\end{tabular}

\vspace{5ex}
\noindent
{\bf Abstract}: 
After a review of theoretical motivations to 
consider theories with
direct couplings of scalar fields to Ricci and gauge curvature
terms, we consider
the dynamics and non--perturbative stabilization of a dilaton
in three and in four dimensions.
In particular, we derive generalized Coulomb potentials
in the presence of a dilaton and discuss a low energy effective
dilaton
potential induced by instanton effects and the S--dual coupling
to axions.
We conclude with a discussion of cosmological implications 
of a light dilaton.

\newpage
\rm
\section{Introduction}

It is a widespread belief both in elementary particle physics and in general
relativity that theoretical Ans\"atze for physics at the Planck scale
are lacking experimental relevance today and in the foreseeable future.
It is indeed a generic feature of theories unifying gravity and quantum field
theory to make predictions mostly for physics at the Planck scale, and eventually
also for
a GUT scale a few orders of magnitude below the Planck scale.
However, many seriously pursued proposals for a framework of quantum
gravity, including string theory, predict 
scalar particles which couple both
to gravity and matter fields. These scalar particles must be very weakly
coupled or have to acquire large
masses in order to meet experimental and cosmological constraints, or there
must exist other non--perturbative mechanisms to make these scalars
invisible in the low energy regime.

A particularly interesting deviation from standard Einstein--Yang--Mills theory
is the prediction of a dilaton. There is no unique definition of dilatons,
but  a common feature of dilatons appearing in different theoretical frameworks
is their direct coupling to Ricci or gauge curvature terms.
A typical feature of the dilaton in string theory 
is its exponential coupling to Yang--Mills terms in the frame where
the dilaton decouples from the curvature scalar $R$, and emphasis
in the present paper will be on a scalar field $\phi$
which couples
to gravity and gauge fields in four dimensions through  
\[
\frac{1}{\sqrt{-g}}{\cal L}=\frac{1}{2\kappa}R-\frac{1}{2}g^{\mu\nu}
\partial_\mu\phi\cdot\partial_\nu\phi
-\frac{1}{4}\exp(\frac{\phi}{f_\phi})
F_{\mu\nu}{}^j F^{\mu\nu}{}_j,
\]
where the mass scale $f_\phi$
characterizes the strength of the coupling. Following usual conventions
we denote the coupling scale $f_\phi$ also as a decay
constant, since in many formulas it appears similar to a decay constant
of a scalar bound state. However, there is an
important difference: The dilatation current containing a 
piece $f_\phi\partial_\mu\phi$
does not parametrize a physical decay amplitude of fundamental
string or Kaluza--Klein dilatons, and $f_\phi$ enters only with negative powers
in actual transition amplitudes.

 In a different setting fundamental scalars are introduced with 
a direct coupling to
the curvature scalar
to mimic a time dependent gravitational constant, to serve as an additional
gravitational degree of freedom,
or for the sake of local scale invariance. 
A scalar 
coupling both to the Ricci scalar and to the Maxwell term
was investigated for a long time by Jordan and his collaborators.
Jordan's motivation originated from Dirac's proposal of a variable gravitational
constant \cite{dirac0}, and from the observation that Kaluza--Klein theory
supports Dirac's proposal if the four--dimensional metric is directly
induced from a five--dimensional metric without rescaling. The coupling of
the scalar to a Maxwell term is then a direct consequence of Kaluza--Klein theory.
However, later Jordan abandoned the coupling to electrodynamics, thus anticipating
a particular case of a Brans--Dicke theory of gravity \cite{pj}.

Motivated from Mach's principle
Brans and Dicke introduced a scalar $\Phi_{\cal BD}$ coupling 
through $R\Phi_{\cal BD}$
to account for a long range scalar field
participating in the gravitational interaction \cite{BrDi}. 
In order to maintain the property that gravitational fields
can be gauged away locally, they required from the outset
ordinary metric couplings of all matter fields, thus excluding in particular
the coupling of the scalar to gauge fields.  
More general scalar--tensor theories of gravity have later been 
defined as theories
which can be related to a Brans--Dicke theory 
through a $\Phi_{\cal BD}$--dependent
rescaling of the metric \cite{rvw}, but 
due to the Weyl invariance of Yang--Mills terms in four dimensions these
theories also contain no direct coupling to the Yang--Mills curvature.

In spite of our emphasis on couplings to Yang--Mills terms,
scalar--tensor theories will be of some interest to us, due to a
theoretical
ambiguity in low energy string theory: It is known since long that the low energy
and low curvature limit of string theory in the critical dimension is given by
Einstein gravity, and this property persists in lower dimensions if low--dimensional
metrics are embedded appropriately in higher--dimensional metrics.
However, the claim that the leading curvature term in four--dimensional string
effective actions is given by the Einstein--Hilbert term
was challenged recently by Gasperini and Veneziano, see \cite{GV,mg} and references
there. A compactification which rescales the low--dimensional metric
by the string dilaton instead of a Kaluza--Klein dilaton
gives a Brans--Dicke type theory in the curvature sector, and the problem
whether low energy gravity in string theory is described by Einstein gravity
in the Einstein frame or by a scalar--tensor theory in the string frame
is an experimental issue. 

The gravitational sector
of a scalar--tensor theory with constant Brans--Dicke parameter
 $\omega_{\cal BD}$ reads
\begin{equation}\label{LBD}
{\cal L_{BD}}=\frac{1}{2}\sqrt{-g}\Phi_{\cal BD}\Big(R-\omega_{\cal BD}g^{\mu\nu}
\partial_\mu\ln(\Phi_{\cal BD})\cdot\partial_\nu\ln(\Phi_{\cal BD})
+2\Lambda(\Phi_{\cal BD})\Big)
\end{equation}
\[
=\sqrt{-g}\Big(\frac{1}{8|\omega_{\cal BD}|}\phi^2R
-\mbox{sgn}(\omega_{\cal BD})\frac{1}{2}g^{\mu\nu}
\partial_\mu\phi\cdot\partial_\nu\phi+\Lambda(\frac{\phi^2}{4|\omega_{\cal BD}|})
\frac{\phi^2}{4|\omega_{\cal BD}|}\Big).
\]

We have to be careful with the sign of $\omega_{\cal BD}$ since the 
Brans--Dicke parameter
can attain negative values. We will see in section \ref{kaka} that a ``strong'' 
version of 
Kaluza--Klein theory yields Brans--Dicke parameters $0>\omega_{\cal BD}>-1$, and 
string theory
in the string frame has $\omega_{\cal BD}=-1$.

A Brans--Dicke type coupling of  a scalar field also emerged
in constructions of locally Weyl invariant theories without
a Weyl vector, see \cite{jw1,bz1,je,gd} and references there.
Coupling 
of a massless scalar $\phi$ to other particle masses 
 through the substitution
\[
m\to m\phi
\]
ensures Weyl invariance with standard kinetic terms
if the rescaling of the metric is accompanied by appropriate rescalings
of scalar and spinor fields. Then local scale invariance
can be ensured to first order if $\partial_\mu\ln\phi$ is used as a connection
and if the kinetic term for $\phi$ corresponds to a
Brans--Dicke theory in the singular limit $\omega_{\cal BD}\to -\frac{3}{2}$.

Both in scalar--tensor theories and in Weyl invariant theories
the dilaton $\phi$ does not couple to Yang--Mills terms.
However, it does couple both to Ricci and gauge curvature terms
in string theory in the string frame and in Kaluza--Klein theory
without metric rescaling, whereas it decouples 
from the Ricci scalar in Weyl transformed Kaluza--Klein theory
and in string theory in the Einstein frame:\\[0.2ex]

\begin{center}
\begin{tabular}{|l|c|c|c|}\hline
Theory & Curvature term & Yang--Mills term & Mass terms\\ \hline
String theory, string frame & $\phi^2 R$  & $\phi^2 F^2$  
&  $m$   \\ \hline
String theory, Einstein frame &  $R$  & $\exp(\phi/f_\phi)F^2$  
&  $m$   \\ \hline
Brans--Dicke theory & $\phi^2 R$  & $F^2$  
&  $m$   \\ \hline
Weyl invariant theory & $\phi^2 R$  & $F^2$  
&  $m\phi$ \\ \hline 
\end{tabular}
\end{center}
\begin{center}
Table 1:\ Theories with  fundamental scalars coupling to curvature terms.\\[0.5ex]
\end{center}

In the normalization of these terms a canonical kinetic term
for the scalar $\phi$ was assumed. The ``strong'' and ``weak'' versions
of Kaluza--Klein theory are related to string theory in the string frame
and in the Einstein frame, respectively.
In neglecting mass dependent couplings of the dilaton in the string
and Kaluza--Klein framework we assumed that the mass terms originate
at a scale far below any string or compactification scale. Otherwise
couplings to mass terms would also appear in this sector, see section \ref{kaka}.

Generically different rows in this table can be connected through field dependent
rescalings of the metric. This can be a useful mathematical tool
in analyzing e.g.\ equations of motion. However, it must be stressed that
field dependent
rescalings of metrics
are not symmetry transformations of one and the same physical system,
but generically have different 
physical implications.

It must also
be emphasized that Table 1 is by no means exhaustive: 
A theory may contain several dilatonic degrees of freedom, and
a particularly important example is compactified
heterotic string theory: This contains besides the model independent 
string dilaton $\phi_S$
at least one Kaluza--Klein dilaton $\phi_K$, and a linear combination $\phi$ of the two
couples to four--dimensional gauge fields, 
whereas in non--rescaled compactification or in the string frame $\phi_K$ or $\phi$
couple to $R$, respectively, while no dilaton couples to $R$ in the Einstein frame.
 Furthermore, in recent years string theory motivated discussions 
of more complicated coupling functions both in the curvature and in the
Yang--Mills sector.

Couplings of light scalars to Ricci or Yang--Mills curvature have very interesting
cosmological implications, and in particular a direct coupling of a scalar to the
Einstein--Hilbert term can have drastic consequences like removing
the initial singularity of space--time without a string threshold.
Cosmological implications of a Brans--Dicke scalar with and without
a cosmological function $\Lambda$ were discussed in \cite{cw2}.
String theory in the string frame has a dilaton with a Brans--Dicke type coupling,
and the work of Veneziano and his collaborators attracted much interest in 
the resulting Brans--Dicke type
cosmology known as string cosmology \cite{GV}. The cosmological implications
 of a dilaton
in Weyl invariant theories 
were investigated in \cite{PSW,BuDr}.
 For a disussion of a low energy string dilaton in the Einstein frame, see e.g.\
 \cite{rdhei,rdmpla3}. There a duality invariant coupling between the dilaton and
the axion played a crucial role in identifying a mass generating mechanism
for the dilaton. However, a possible cosmological significance of axion--dilaton
interactions was emphasized already in \cite{wb}, where the common appearance
of axions and dilatons was motivated from supersymmetry \cite{Wilog,DIN2}.

Superstring theory currently
undergoes a major change of paradigm:
Numerous duality symmetries have been established or proposed between
seemingly different versions of the theory in various macroscopic dimensions,
and it has become particularly clear that symmetry transformations 
interchanging
axions and dilatons play a significant and fundamental role in the theory.
These symmetries go by the name strong/weak coupling duality, or S--duality
for short, because they involve sign flips of the dilaton.
Since the expectation value of the exponentiated dilaton determines the
strength of the string coupling, those sign
flips can interchange strongly coupled  regimes of string theory
with weak coupling regimes. 

We will see that under a certain constraint on decay 
constants of the axion and the dilaton to be
explained in section \ref{stab4d},
a low energy imprint of S--duality can still mix the dilaton and
the axion.
This indicates in particular that a fundamental light axion should come with
a dilaton, and that their properties should be intimitely connected.
Now it is known since long that the gauge theory of strong 
interactions \cite{FGM} 
strongly
motivates considerations of a light axion, and that instantons create an effective
potential for this axion.
If the QCD axion is a fundamental pseudo--scalar, and if axion--dilaton
duality is realized at some scale, then we should also expect a fundamental dilaton
coupling to QCD. The main motivation for the present work was the
observation that instantons induce an interesting
potential for such a dilaton far below the scale of supersymmetry breaking.
Such a possibility was not pursued before, since an
estimate on the coupling of a QCD dilaton to nucleons seems to violate the
Newton approximation for weak gravitational fields, see \cite{EKOW} and references there:

Constraints on dilaton masses without a direct coupling to the Ricci scalar
are based on the assumption
that the dilaton can be treated as a local gauge coupling which shows up
in nucleon masses. This might imply a coupling of the dilaton to hadronic mass
terms in low energy effective theories, eventually also
inducing material dependent couplings to
macroscopic bodies.
An estimate derived on this basis states
that the dilaton should be heavier than $10^{-4}$ eV
to be on the safe side, too heavy for
a QCD dilaton with a coupling scale $f_\phi$ of the order of the 
Planck mass\footnote{Our conventions for Planck units are
$m_{Pl}=(8\pi G)^{-1/2}=2.4\times10^{18}$ GeV, $t_{Pl}=2.7\times 10^{-43}$ s,
$l_{Pl}=8.1\times 10^{-35}$ m.}.
The reasoning implied in the derivation
of this estimate was ingenious, but a weak point concerns
the interpretation of the dilaton as a local gauge coupling: 
We will calculate the impact of a dilaton on the classical $1/r$
interaction of gauge charges in section \ref{copot} and find that
the dilaton either regularizes the Coulomb potential 
at a distance $r_\phi\sim f_\phi^{-1}$
or implies confinement, with an interaction potential between stationary
charges raising linearly with the distance. 
These results indicate that the impact
of a dilaton in gauge theory
is not adequately approximated by an effective local gauge
coupling, and it seems 
premature to conclude that the dilaton--gluon coupling induces a dilaton--nucleon
coupling proportional to the mass terms
in low energy hadron theories. 
The problem in how far such a coupling
would amount to material dependent couplings
also causes some uncertainty, and
lacking a reliable quantitative
picture of the emergence
of the hadron spectrum in QCD both 
with and without a dilaton, we are currently unable
to discuss constraints from the weak
equivalence principle on macroscopic bodies. 
On the other hand, constraints
from elementary particle masses in the standard model do not arise,
since the dilaton originates at string or Kaluza--Klein scales far beyond the
weak scale, and therefore neither the string dilaton nor
a Kaluza--Klein dilaton are expected to couple to mass terms
in the standard model.
However, in section \ref{cosmo} we will also discuss implications of
a dilaton with a mass $m_\phi\geq 10^{-4}$ eV. If the mechanism
of axion induced dilaton stabilization outlined in section
\ref{stab4d} is correct this mass would correspond to a decay 
constant $f_\phi\leq 10^{11}$ GeV
far below the Planck scale, but it would still be invisible
from a particle physics point of view and have
interesting cosmological implications. Of course, from a stringy
point of view such a low decay constant is a puzzle, since string theory
generically predicts dilaton coupling scales of the order of the 
Planck mass.

While the results of section \ref{copot} show that the dilaton in gauge theory
can not be addressed as an effective local gauge coupling, the 
expectation value of the dilaton still
seems to imply an ambiguity in the definition of the coupling.
To elucidate this consider the Lagrangian (in flat Minkowski space)
\[
{\cal L}=-\frac{1}{2}\partial_\mu\phi\cdot\partial^\mu\phi
-\frac{1}{4}\exp(\frac{\phi}{f_\phi})
F_{\mu\nu}{}^j F^{\mu\nu}{}_j 
+\overline{\psi}(i\gamma^\mu\partial_\mu+q\gamma^\mu A_\mu-m)\psi
\]
describing gauge theory with a dilaton $\phi$, gauge 
coupling $q$ and fermions $\psi$.
 For fixed $\psi$ the mapping 
\[
q\to q'
\]
\[
A_\mu\to {A'}_\mu=\frac{q}{q'}A_\mu
\]
\[
\phi\to\phi'=\phi-2f_\phi\ln(\frac{q}{q'})
\]
leaves the Lagrangian invariant but implies e.g.\ a rescaling of 
the 1--gluon exchange
4--fermion
amplitude ${\cal A}_{2\to 2}\to \frac{{q'}^2}{q^2}{\cal A}_{2\to 2}$. 
This is the gauge theory
version of the problem of the running dilaton or of the constancy of constants 
(see e.g.\ \cite{DS}):
Motion of the expectation value of the dilaton implies a rescaling of the 
gluon propagator
and a corresponding rescaling of the gauge coupling measured in scattering events.

On the other hand, at least the variation
of the electromagnetic fine structure constant $\alpha_{em}=\frac{e^2}{4\pi}$
over cosmic time scales is strongly constrained:
Improving on an analysis by Shlyakhter \cite{ais}, Damour and Dyson point out
that the fine structure constant about
two billion years ago, when the natural Oklo fission reactor in West Africa 
was active,
differs from todays value at 
most by $\frac{|\delta\alpha|}{\alpha}< 1.2\times 10^{-7}$ \cite{DD}.
On larger time scales, Varshalovich et al.\ 
give $\frac{|\delta\alpha|}{\alpha}< 1.6\times 10^{-4}$ for the variation of 
the fine structure
constant between ultraviolet emission of high 
red shift ($z\sim 3$) quasars and today \cite{VPI}.
In a flat universe electromagnetic waves emitted at
 $z= 3$ travelled for almost 88\% of the lifetime of the universe
(i.e.\ they were emitted certainly more than 9 billion years ago), and if a 
dilaton also
couples to the photon this means that some mechanism must have stabilized the
expectation value of the dilaton at an early stage
in the evolution of the universe.

Usual attempts to solve the stability problem
for the dilaton link the generation of a low energy
dilaton potential to gaugino condensation: Due to the dilaton gaugino
coupling implied by supersymmetry, a gaugino condensate provides an
attractive mechanism to generate a dilaton mass. However, it must be
pointed out that a gaugino
condensate in the most direct and simple supersymmetric Yang--Mills
dilaton theory would send the dilaton expectation value to $-\infty$
rather than stabilizing it at some finite value. This is a simple consequence
of the fact
that the dilaton couples to gluinos with an exponential $\exp(2\phi/f_\phi)$,
i.e.\ with the same sign as in the dilaton gluon coupling. Therefore anomalous
low energy effective terms \cite{VY,trt} or duality invariant potentials have 
to be
employed to stabilize dilatonic degrees of freedom, see e.g.\ \cite{FILQ1,hpn}
and references there.

The puzzle can be resolved in a different way
by noting that the dilaton couples with a term
 $\exp(-2\phi/f_\phi)$ to the kinetic energy of the axion, whence
a non--vanishing variance of the axion would provide a direct and
simple way to generate a dilaton mass \cite{rdhei,rdmpla3}. This mechanism 
works with or without a dilaton gluino coupling, and the discussion
in section \ref{stab4d} concentrates on the non--supersymmetric case.

In a string inspired four--dimensional field theory the dilaton coupling to gauge
fields will generically be a
superposition of the massless closed string excitation accompanying
the graviton and a Kaluza--Klein dilaton defined as a logarithm
of the determinant
of the internal metric. In compactifications of heterotic string theory, e.g., 
the four--dimensional dilaton is dominated by a Kaluza--Klein component
with
a mixing angle of $30^0$ into the string dilaton.

Like the string dilaton the Kaluza--Klein dilaton comes initially without
a mass.
This initial absence of a dilaton potential in the low energy sector
is easy to understand: Since the low energy effective fields are zero modes
of the internal derivative operators, they couple to the fluctuations
of the internal dimensions, but carry no remembrance of their actual
size. Therefore, we may expect dilaton stabilization in the low energy sector
only if there exist non--perturbative effects which are genuinely related
to the number of macroscopic dimensions, since these effects would have
to disappear
upon decompactification or further compactification. We know
such a genuinely four--dimensional non--perturbative effect very
well: Instantons in gauge theories may provide a mechanism to stabilize four
dimensions, i.e.\ we suspect that instantons induce an effective
dilaton potential in the low energy regime. However, suppose we start
with some large gauge group $G$ which decomposes into several
abelian and non--abelian factors: Where do we expect the dominant
instanton contributions? Instanton energies go with the inverse gauge coupling
squared, but instantons are suppressed in broken gauge groups due to the
large masses of the gauge fields. Therefore the dominant
instanton contributions should arise in that factor of $G$ which corresponds
to the non--abelian unbroken symmetry with the largest coupling constant.
If instantons stabilize the dilaton, it is QCD which has to make the dominant
contribution!

Of course, there may appear other non--perturbative effects in four dimensions,
eventually breaking supersymmetry and active at a much higher scale.
It is well conceivable that such effects may also create an effective 
dilaton potential,
and in a sense this is a working hypothesis for mainstream research in the field.
I make no attempt to invalidate the mainstream approach which links the
dilaton potential to supersymmetry breaking, but I suppose there is  
enough motivation to consider an axion--dilaton system which
is stabilized at or above the QCD phase transition, around $10^{-4}$ 
seconds after the big bang.

We have defined the dilaton in four dimensions through its characteristic
coupling to gauge fields, yet we have to re--address its coupling to gravity,
in order to explain why we exclude such couplings in the present paper:
The dilaton can couple to the Einstein--Hilbert term through a term $U(\phi)R$,
and we have pointed out already
that direct compactification of dimensions would predict a polynomial
coupling of Kaluza--Klein dilatons both to $R$ and to gauge fields.
In the gravitational sector this would correspond to a
Brans--Dicke type theory of gravity 
with a 
Brans--Dicke parameter $\omega_{\cal BD}\simeq -1$ (see section \ref{kaka}). 
However,
through appropriate Weyl rescalings the dilaton
can be arranged to couple to gravity in a standard way without coupling
directly to the curvature scalar $R$. The set of fields and the metric where
the lowest order graviational action is given by the 
Einstein--Hilbert term $\int d^4x\sqrt{-g}R$ is the Einstein frame, and 
we will mainly use this frame.
This is not just a matter of taste, but chosen on the basis of
experimental constraints:
Solar system measurements of light bending
and time delay in gravitational fields are known to constrain 
deviations from standard
Einstein gravity to less than $10^{-3}$, and this imposes rather strong limits on 
scalar--tensor
theories of gravity. It implies in particular that the inverse
Brans--Dicke parameter
measuring the direct coupling of light
scalar fields to curvature is very small,
implying in turn that even if the physical metric would not correspond to
an Einstein frame, it would have to be very close to an Einstein frame.
It was also mentioned before that the string dilaton does not couple to the
Einstein--Hilbert term in the critical dimensions 10 and 26 \cite{GSW,BH,LT}.

There are other viable alternatives than directly relying on a Weyl
transformed metric as the physical metric: If the Brans--Dicke scalar
somehow
acquires such a large mass that it effectively is frozen to a constant
value, this would certainly comply with all contemporary tests of Einstein
gravity. Furthermore, Damour and Nordtvedt pointed out that
a scalar--tensor theory of gravity with a convex coupling function $U(\phi)$
also effectively reduces to Einstein gravity, even without a mass term \cite{DN}.

In order to faciliate the comparison with calculations in other frames, 
and to clarify
which models are related through Weyl rescalings, it is useful to have 
a dictionary of the behavior of various sectors in the Lagrangian under Weyl 
transformations:
Under rescalings of the metric in $D$ dimensions
\[
\tilde{g}_{\mu\nu}=
\exp\!\Big(2\frac{\lambda-\tilde{\lambda}}{D-2}\phi\Big)g_{\mu\nu}
\]
the gravitational sector transforms according to
\[
\sqrt{-\tilde{g}}\tilde{g}^{\mu\nu}\exp(\tilde{\lambda}\phi)
\big(\tilde{R}_{\mu\nu}+\frac{D-1}{D-2}\tilde{\lambda}^2
\partial_\mu\phi\cdot\partial_\nu\phi\big)
=\sqrt{-g}g^{\mu\nu}\exp(\lambda\phi)\big(R_{\mu\nu}+\frac{D-1}{D-2}\lambda^2
\partial_\mu\phi\cdot\partial_\nu\phi\big).
\] 
In the matter sector one finds for Yang--Mills fields
\[
\sqrt{-\tilde{g}}\tilde{g}^{\mu\nu}\tilde{g}^{\rho\sigma}\exp(\tilde{\alpha}\phi)
\mbox{tr}(F_{\mu\rho}\cdot F_{\nu\sigma})=
\sqrt{-g}g^{\mu\nu}g^{\rho\sigma}\exp(\alpha\phi)
\mbox{tr}(F_{\mu\rho}\cdot 
F_{\nu\sigma})
\]
with
\[
\alpha=\tilde{\alpha}+\frac{D-4}{D-2}(\lambda-\tilde{\lambda}),
\]
while fermions coupling to Yang--Mills fields $A$ and a canonical spin connection
$\Omega$ transform according to
\[
\sqrt{-\tilde{g}}\bar{\tilde{\psi}}
[\tilde{e}^\mu{}_a\gamma^a(\partial_\mu+\tilde{\Omega}_\mu
-iqA_\mu)+im]\tilde{\psi}
=\sqrt{-g}\bar{\psi}[e^\mu{}_a\gamma^a(\partial_\mu+\Omega_\mu-iqA_\mu)
+\exp\!\Big(\frac{\lambda-\tilde{\lambda}}{D-2}\phi\Big)im]\psi
\]
with
\[
\tilde{\psi}=
\exp\!\Big(-\frac{D-1}{D-2}\frac{\lambda-\tilde{\lambda}}{2}\phi\Big)\psi,
\]
\[
\tilde{\Omega}^a{}_{b\mu}=\Omega^a{}_{b\mu}+\frac{\lambda-\tilde{\lambda}}{D-2}
(e_\mu{}^a e^\nu{}_b-e_{\mu b}e^{\nu a})\partial_\nu\phi.
\]

As an application, the combined Weyl
transformation and redefinition
\[
g_{\mu\nu}=F(\Phi)^{\frac{2}{D-2}}\tilde{g}_{\mu\nu}
\]
\[
\phi=
\int d\Phi
\sqrt{\frac{G(\Phi)}{F(\Phi)}+\frac{D-1}{(D-2)\kappa}\frac{F'(\Phi)^2}{F(\Phi)^2}}
\]
transform Jordan--Brans--Dicke type 
theories into the (physically inequivalent) Einstein frame:
\[
\sqrt{-\tilde{g}}\tilde{g}^{\mu\nu}
\big(\frac{1}{2\kappa}F(\Phi)\tilde{R}_{\mu\nu}-\frac{1}{2}
G(\Phi)\partial_\mu\Phi\cdot\partial_\nu\Phi\big)=
\sqrt{-g}g^{\mu\nu}\big(\frac{1}{2\kappa}R_{\mu\nu}-\frac{1}{2}
\partial_\mu\phi\cdot\partial_\nu\phi\big),
\]
and according to the previous paragraph we may also transform the kinetic terms
of the fermions to standard form.

Our analysis in four dimensions will be mainly based on gauge theory 
coupled to an axion $a$ and a dilaton $\phi$ in the Einstein frame:
\begin{equation}\label{lag1}
\frac{1}{\sqrt{-g}}{\cal L}=\frac{1}{2\kappa}R-\frac{1}{2}g^{\mu\nu}
\partial_\mu\phi\cdot\partial_\nu\phi
-\frac{1}{2}\exp(-2\frac{\phi}{f_\phi})g^{\mu\nu}
\partial_\mu a\cdot\partial_\nu a
\end{equation}
\[
-\frac{1}{4}\exp(\frac{\phi}{f_\phi})
F_{\mu\nu}{}^j F^{\mu\nu}{}_j +
\frac{q^2}{64\pi^2 f_a}\epsilon^{\mu\nu\rho\sigma} a F_{\mu\nu}{}^j
F_{\rho\sigma j}.
\]

Superficially, we will refer to this system as an axion--dilaton--gluon
system, irrespective of whether the gauge group is SU(3) or some other
Lie group.

The particular ratio between the couplings of the dilaton to
gauge fields and the axion in (\ref{lag1}) can be motivated for two reasons:
On the one hand, this ratio of couplings arises automatically
in Kaluza--Klein compactifications to four dimensions, as will be
elucidated in section \ref{kaka}, while on the other hand the same 
ratio also arises from the requirement of 
duality symmetry
in the axion--dilaton system.
The last, and maybe most important motivation for consideration
of the axion--dilaton--gluon system comes from the observation that
this system must hold a clue to the issue of dilaton stabilization
in four dimensions: It is known
that instantons create an effective axion potential $V(a)\sim -\cos(a/f)$
plus higher order terms in $\cos(a/f)$.
We will see that instantons and the effective axion potential
survive introduction of the dilaton and break the scale invariance
of the equations of motion following from (\ref{lag1}).
This provides a strong hint that instantons and axions must also
lift the degeneracy of the dilaton potential associated with the scale
invariance of the tree level equations of motion. 

The resulting dilaton potential has many interesting implications on the low
energy sector of string theory  and cosmology.
In particular, the dynamics of the
dilaton switches from expansion dominance to an oscillatory behavior
around $10^{-4}$ seconds after the initial singularity, that is
``long'' after the onset of axion oscillations ($\sim 10^{-6}$ seconds).
At this time the temperature of the universe has already
dropped to a value
near the QCD phase transition, and
the dilaton can make
an appreciable contribution to the energy density of the universe
as a cold dark matter candidate if its variance above 1 TeV
is of the order of the Planck mass $\sqrt{\phi^2}\sim m_{Pl}$.

Investigation of the dilaton potential resulting from the observations
outlined above and discussions of cosmological implications 
will be a primary concern in the present work, and we will concentrate
on these tasks especially in sections \ref{stab4d}--\ref{cosmo}.
Besides this, we will discuss stabilization of the dilaton in three
dimensions in section \ref{dil3d}, while the impact of the dilaton
on potentials of pointlike particles in four--dimensional gauge theory 
is examined
in section \ref{copot}.
Sections \ref{inst} and \ref{kaka} survey introductory material needed in the 
discussion
of the dilaton potential. While no attempt was made to make the paper
fully self--contained, the introductory
sections should make it amenable
to non--experts with some basic knowledge in quantum field theory, string theory
and cosmology.

\newpage
\section{Instantons in gauge theory}\label{inst}

Instantons are classical solutions of Yang--Mills equations on Euclidean
four--dimen\-sio\-nal manifolds \cite{BPST,tH1,ADHM}. 
They play a prominent role in particle physics through their contribution
to 't Hooft's solution of the strong U(1) problem and through their contribution
to chiral symmetry breaking:
Chiral symmetry breaking is signaled through a quark condensate
which can be related to the spectral density $\rho(\lambda)$ of the 
Euclidean Dirac
operator through a celebrated relation of Banks and Casher \cite{BC}
\[
\langle\overline{q}q\rangle =-\pi\rho(0).
\]
While we are still lacking a full understanding of
the dynamics of chiral symmetry
breaking in QCD from first principles,
the emergence of a non--vanishing spectral density
at eigenvalue zero can be
attributed to the presence of an instanton liquid in the Euclidean
vacuum, as has been thoroughly reviewed in \cite{ScSh,avs}.

In the present setting we are interested in
instantons because some appropriate generalization of them will contribute
to the effective potential of the dilaton, and we will review some
of the properties of instantons in this section. Classical and very 
thorough reviews
of instantons have been given in \cite{sc,VZNS}. For the ADHM 
construction of the general multi--instanton
solution see \cite{ADHM} and \cite{CG} and references there.
The path breaking work on the calculation of quantum
effects was \cite{tH1}.

In this introductory section we will only review the one--instanton
solution elaborating on a theorem
of Wilczek. Wilczek's theorem relates instantons to conformally flat 
spaces of constant scalar
curvature \cite{fw1,BS,rdlmp1} and has the virtue to naturally 
yield 't Hooft's Ansatz and to explain
the group theoretic origin of the 't Hooft symbols \cite{tH1}.

Throughout this section we will work with two conformally
related Euclidean 4--spaces, one of those being flat while the 
other metric can
be written as
\[
g_{\mu\nu}(x)=\chi^2(x)\delta_{\mu\nu}=
\frac{1}{\sigma^2(x)}\delta_{\mu\nu}.
\]
Indices raised with $g^{\mu\nu}(x)$ will be denoted by a dot in this 
section: $g^{\mu\nu}(x)v_{\nu}(x)=v^{\dot{\mu}}(x)$. The gauge 
covariant derivative
is\footnote{We are using an anti--hermitian basis for ${\cal A}$. 
The relation between
the gauge potential ${\cal A}_{\mu}\equiv {\cal A}_{\mu i}(-iX^i)$ 
and $A_\mu\equiv A_{\mu i} X^i$
is ${\cal A}_{\mu i}=qA_{\mu i}$.} $D_\mu=\partial_\mu+{\cal A}_\mu$ 
and the covariant derivative in the conformally
flat space is $\nabla_\mu$. Since we will not perform Legendre 
transformations in the Euclidean setting
and the Euclidean partition function resembles a canonical 
ensemble, we will use the terms action
and energy synonymously in this section.

The well-known local isomorphism
SO$(4)\cong \mbox{SU}(2)\times \mbox{SU}(2)$ may be used to 
reduce gl$(4,\mathbb{R})$ 
connections $\Gamma$
to su$(2)$ connections ${}^{\pm}{\cal A}$ according to
\begin{equation}\label{reduct}
{}^{\pm}{\cal A}_{\mu i}=
-(^{\pm}Z_i)^\alpha{}_\beta \Gamma^\beta{}_{\alpha\mu},
\end{equation}
where an explicit representation of the SU$(2)$ projectors is
\[
(^{\pm}Z_i)_{\mu\nu}=
\frac{1}{2}(\pm \delta^0{}_\nu \delta_{i\mu}\mp \delta^0{}_\mu \delta_{i\nu}-
\epsilon_{ijk}\delta^j{}_\mu \delta^k{}_\nu).
\]
These projectors provide a selfdual and an anti--selfdual 
basis for four--dimensional 
representations of su(2), and in compliance with the
uniqueness of the spin--$\frac{3}{2}$ equivalence class, they are intertwined 
via $\mathbb{T}\cdot {}^+Z_i\cdot\mathbb{T}={}^-Z_i$ 
with $\mathbb{T}=\mbox{diag}\{-1,1,1,1\}$.
Identifying indices which carry the values 0 or 4, these
generators are related to the 't Hooft symbols via
\[
(^+Z_i)_{\mu\nu}= 
-\frac{1}{2}\overline{\eta}_{i\mu\nu},\qquad (^-Z_i)_{\mu\nu}= 
-\frac{1}{2}\eta_{i\mu\nu}.
\]
Thus the appendix of \cite{tH1} may be used if some signs are adjusted properly
due to $\epsilon_{0123}=-\epsilon_{1234}$. The reduction
according to (\ref{reduct}) yields su$(2)$--valued curvatures
\begin{equation}
{}^{\pm}{\cal F}_{\mu\nu}{}^i=
-(^{\pm}Z^i)^\alpha{}_\beta (R^\beta{}_{\alpha\mu\nu} +S^\beta{}_{\alpha\mu\nu})
\end{equation}
with the deviation from the Riemannian
\[
S^\alpha{}_{\beta\mu\nu} = 
\frac{1}{2}\Gamma^\rho{}_{\beta\mu}(\Gamma^\alpha{}_{\rho\nu}
+\Gamma_\rho{}^\alpha{}_{\nu})-
\frac{1}{2}\Gamma^\alpha{}_{\rho\mu}(\Gamma^\rho{}_{\beta\nu}
+\Gamma_\beta{}^\rho{}_{\nu}).
\]
The conformal  Ansatz is now introduced by inserting the Levi--Civit\'{a} connection
\begin{equation}
\Gamma^\rho{}_{\mu\nu}=
\frac{1}{2\sigma^2}(\delta_{\mu\nu}\partial^\rho\sigma^2-\delta^\rho{}_\mu
\partial_\nu\sigma^2-\delta^\rho{}_\nu\partial_\mu\sigma^2)
\end{equation}
corresponding to the conformally flat metric $g$.
This Ansatz leads to $S=0$, and (\ref{reduct}) reduces to the
't Hooft Ansatz:
\begin{equation}
{}^{\pm}{\cal A}_{\mu i}=(^{\pm}Z_i)^\rho{}_\mu\partial_\rho\ln(\sigma^2).
\end{equation}
The combination of ${}^{\pm}{\cal A}_{\mu}{}^i$ and $\Gamma$ into a generally covariant
derivative: 
\begin{equation}
{\cal D}_{\mu}{}^{\pm}{\cal F}_{\lambda\nu}=\partial_\mu {}^{\pm}{\cal F}_{\lambda\nu}+
[{}^{\pm}{\cal A}_{\mu},{}^{\pm}{\cal F}_{\lambda\nu}]
-{}^{\pm}{\cal F}_{\lambda\sigma}\Gamma^{\sigma}{}_{\nu\mu}-
{}^{\pm}{\cal F}_{\sigma\nu}\Gamma^{\sigma}{}_{\lambda\mu}
\end{equation}
leaves the su(2)--projectors invariant:
\[
{\cal D}_{\mu}(^{\pm}Z_i)_{\lambda\nu}=
\nabla_{\mu}(^{\pm}Z_i)_{\lambda\nu}+
\epsilon_{ijk}{}^{\pm}{\cal A}_{\mu}{}^j (^{\pm}Z^k)_{\lambda\nu}=0.
\]
As a consequence covariant differentiation and su(2) reduction
commute:
\begin{equation}\label{DD1}
{\cal D}_{\mu}{}^{\pm}{\cal F}_{\lambda\nu}=
-(^{\pm}Z_i)^{\alpha}{}_\beta\nabla_\mu
R^\beta{}_{\alpha\lambda\nu}.
\end{equation}
Furthermore, the gauge covariant divergence and the generally covariant
divergence of the field strength are related in a simple way:
\begin{equation}\label{DD2}
D_\mu{}^{\pm}{\cal F}^{\mu}{}_{\nu i}=
\frac{1}{\sigma^2}{\cal D}_{\mu}{}^{\pm}{\cal F}^{\dot{\mu}}{}_{\nu i}.
\end{equation}

Equations (\ref{DD1},\ref{DD2}) imply a simple relation 
between the divergences of $\cal F$
and $R$:
\begin{equation}\label{DD3}
D_\mu{}^{\pm}{\cal F}^{\mu}{}_{\nu i}=
-\frac{1}{\sigma^2}(^{\pm}Z_i)^\alpha{}_\beta \nabla^{\dot{\mu}}R^\beta{}_{\alpha\mu\nu}
\end{equation}
and this yields a translation of the Yang--Mills equations into a condition on the curvature
of the conformally flat space.
The vanishing of the Weyl tensor implies that
the covariant divergence of the Riemannian in a conformally flat space is 
\[
\nabla^{\dot{\mu}}R^\beta{}_{\alpha\mu\nu}=-\frac{1}{6}(\delta^\beta{}_\nu\partial_\alpha
-\delta_{\alpha\nu}\partial^\beta)R
\]
with the curvature scalar $R=R^{\dot{\mu}}{}_\mu$.
Therefore the gauge covariant divergence of the su(2) curvatures from the
conformally flat space are simply proportional to the gradient of the curvature
scalar:
\begin{equation}\label{DD4}
D_\mu{}^{\pm}{\cal F}^{\mu}{}_{\nu i}(x)=
\frac{1}{3\sigma^2(x)}(^{\pm}Z_i)^\mu{}_\nu\partial_\mu R(x)
\end{equation}
and this expresses Wilczek's theorem \cite{fw1}:
su(2) reduction of the Riemannian connection
in a conformally flat 4--space yields solutions to the Yang--Mills equations
if and only if the curvature scalar is constant.

If the curvature scalar is expressed in terms of the conformal factor $\chi$
this theorem establishes a connection between Yang--Mills theory
and $\chi^4$ theory:
\begin{equation}\label{DD5}
\partial_\mu\partial^\mu\chi+\frac{1}{6}R\chi^3=0.
\end{equation}
However, the formulation in terms of the inverse conformal factor
serves our purposes better:
\begin{equation}\label{DD6}
\partial_\mu\partial^\mu\sigma^2-\frac{3}{2\sigma^2}
\partial_\mu\sigma^2\cdot\partial^\mu\sigma^2-\frac{1}{3}R=0.
\end{equation}
We can not solve this equation in a general fashion. 
However, it was observed in \cite{rdlmp1} that the stronger
condition of local symmetry 
\[
\nabla_\lambda R^{\dot{\mu}}{}_\nu=0
\]
allows for a general solution. In terms of $\sigma(x)$ the condition
of local symmetry takes the following form:
\begin{equation}\label{DD7}
\partial^\mu\partial_\nu\partial_\lambda\sigma^2+\delta^\mu{}_\nu
\partial_\lambda[\sigma\partial_\alpha\partial^\alpha\sigma
-2(\partial_\alpha\sigma)(\partial^\alpha\sigma)]
-[\delta^\mu{}_\nu\partial_\lambda+\delta^\mu{}_\lambda\partial_\nu+
\delta_{\lambda\nu}\partial^\mu](\partial_\alpha\sigma)(\partial^\alpha\sigma)=0.
\end{equation}

Summation over $\mu$ and $\nu$ leads again to
\begin{equation}\label{DD8}
\partial_\lambda R=6\partial_\lambda[\sigma\partial_\alpha\partial^\alpha\sigma
-2(\partial_\alpha\sigma)(\partial^\alpha\sigma)]=
3\partial_\lambda[\partial_\alpha\partial^\alpha\sigma^2
-6(\partial_\alpha\sigma)(\partial^\alpha\sigma)]=0
\end{equation}
and this shows that equation (\ref{DD7}) is equivalent to a set of
equations consisting of (\ref{DD6}) and
\begin{equation}\label{DD9}
\partial_\mu\partial_\nu\partial_\lambda\sigma^2
-\frac{1}{6}(\delta_{\mu\nu}\partial_\lambda+\delta_{\mu\lambda}\partial_\nu+
\delta_{\lambda\nu}\partial_\mu)\partial_\alpha\partial^\alpha\sigma^2=0.
\end{equation}

Eq.\ (\ref{DD9}) can be solved by standard methods, and the simplest method of solution
assuming differentiability to fifth order proceeds as follows:

Contracting (\ref{DD9}) with $\partial^\lambda$ and with $\partial^\lambda\partial^\nu$
yields
\[
\partial_\mu\partial_\nu\partial_\lambda\partial^\lambda\sigma^2=\frac{1}{4}\delta_{\mu\nu}
(\partial_\lambda\partial^\lambda)^2\sigma^2
\]
\[
\partial_\mu(\partial_\lambda\partial^\lambda)^2\sigma^2=0
\]
implying
\[
\partial_\mu\partial_\nu\partial_\rho\partial^\rho\sigma^2=48\lambda^2\delta_{\mu\nu}
\]
where in writing the arbitrary integration constant as a square we already took into
account that our solution in the end should correspond to a definite function.
The previous equation is readily integrated to
\[
\partial_\nu\partial_\lambda\partial^\lambda\sigma^2=48\lambda^2 x_\nu+12 c_\nu
\]
whence (\ref{DD9}) reduces to
\[
\partial_\mu\partial_\nu\partial_\rho\sigma^2=
\delta_{\mu\nu}(8\lambda^2x_\rho+2c_\rho)
+\delta_{\mu\rho}(8\lambda^2x_\nu+2c_\nu)+
\delta_{\rho\nu}(8\lambda^2x_\mu+2c_\mu).
\]
This is readily integrated again and the general solution contains 20 parameters:
\begin{equation}\label{DD10}
\sigma^2(x)=\lambda^2 r^4+c_\mu x^\mu r^2+\frac{1}{2}a_{\mu\nu}x^\mu x^\nu +b_\mu x^\mu
+\zeta^2.
\end{equation}
We may identify the 4--vector $c$ and the off--diagonal components of $a$ as Poincar\'{e}
degrees of freedom, and gauge them away through an appropriate Poincar\'{e}
transformation. 
We then write $a_{\mu\nu}=\Lambda_\mu\delta_{\mu\nu}$ and use the abbreviation
$A\equiv \Sigma_{\mu}\Lambda_\mu-\frac{R}{3}$.
In this gauge equation (\ref{DD6}) translates into the following set of
algebraic constraints on the coefficients in (\ref{DD10}):
\begin{eqnarray*}
A\lambda^2&=&0,\\
2A\zeta^2&=&3b_\mu b^\mu,\\
\lambda^2 b_\mu&=&0,\\
(3\Lambda_\mu-A)b_\mu&=&0,\\
(3\Lambda_\mu-A)\Lambda_\mu&=&48\lambda^2\zeta^2.
\end{eqnarray*}

In case $\lambda^2=0$ there exist only singular solutions with
either a pointlike singularity (corresponding to the meron solution
of De Alfaro, Fubini and Furlan \cite{DAFF}), singular lines, planes or 3--spaces.
In case $\lambda^2>0$ there exist singular solutions with either two
pointlike singularities (2--meron solution), a singular circle, a singular
2--sphere, or a singular 3--sphere. However, there exists one regular solution
given by
\begin{equation}\label{DDinst}
\sigma(x)=\lambda (r^2+\varrho^2).
\end{equation}
The corresponding metric describes a 4--sphere of radius $r_4=\frac{1}{2\lambda\varrho}$
centered at a point $(\varrho-r_4)\vec{e}_5$ in $\mathbb{R}^5$
in stereographic coordinates. The BPST instanton ${}^+{\cal A}$ and 
anti--instanton ${}^-{\cal A}$ are 
\begin{equation}\label{DD11}
{}^\pm{\cal A}_{\mu i}=\frac{2}{r^2+\varrho^2}(\pm\delta^0{}_\mu x_i \mp\delta_{i\mu} x^0
+\epsilon_{ijk}\delta^j{}_\mu x^k)
\end{equation}
and they satisfy
\[
\lim_{r\to\infty}{}^+{\cal A}_{\mu i}(x)\frac{\sigma^i}{2i}=U^{-1}(x)\partial_\mu U(x),
\qquad \lim_{r\to\infty}{}^-{\cal A}_{\mu i}(x)\frac{\sigma^i}{2i}=U(x)\partial_\mu U^{-1}(x),
\]
with
\[
U(x)=\frac{1}{r}(x^0+x^i\sigma_i).
\]
Writing the angle to $\vec{e}_0$, $\cos\vartheta=\frac{x^0}{r}$, it is apparent
that 
\[
U^{\pm n}(x)=\cos(n\vartheta)\pm i\frac{x^i\sigma_i}{\sqrt{x_j x^j}}\sin(n\vartheta)
\]
describe mappings of winding numbers $\pm n$ 
from $S^3$ to SU(2).

The corresponding field strengths are
\begin{equation}\label{DD12}
{}^\pm E_j{}^i=\pm\frac{4\varrho^2}{q(r^2+\varrho^2)^2}\delta_j{}^i,
\qquad
{}^\pm B_j{}^i=-\frac{4\varrho^2}{q(r^2+\varrho^2)^2}\delta_j{}^i,
\end{equation}
and these solutions are apparently (anti--)selfdual under the euclidean duality 
transformation $E\to -B$, $B\to -E$. In the present construction the duality
property arises as
a consequence of the diagonal structure of the field strengths with respect to
space--time and su(2) indices. This ensures that the duality of
 ${}^\pm{\cal F}^\alpha{}_{\beta\mu\nu}
\equiv{}^\pm{\cal F}^i{}_{\mu\nu}(^\pm Z_i)^\alpha{}_{\beta}$
in the internal indices $\alpha$, $\beta$ 
carries over to the space--time indices $\mu$, $\nu$.

The energy density is
\[
{\cal L}=\frac{48\varrho^4}{q^2(r^2+\varrho^2)^4}
\]
yielding an action
\[
S=\frac{8\pi^2}{q^2}.
\]
These solutions and their generalizations to higher winding numbers $n$ appear
also in general SU($N_c$) gauge theory through the various embeddings of
SU(2) subgroups.

It is apparent then that instantons should contribute to the effective dilaton potential
since the dilaton couples directly to $F^2$ and a large dilaton would imply
large instanton energy.

The singular solutions with $\lambda^2\neq 0$ also fall off like $r^{-4}$ at
large distances and therefore really correspond to solutions of the
vacuum Yang--Mills equations, albeit with infinite action.

\newpage
\section{The Kaluza--Klein paradigm}\label{kaka}

Kaluza--Klein theory asserts that part of the scalar and vector fields
and the metric 
in a theory in $d$ dimensions can be identified as components of a 
higher--dimensional
metric, and that the appropriate Lagrangian in $d$ dimensions
can be inferred from a higher--dimensional theory containing additional
compact dimensions. One might distinguish a strong and a weak version
of Kaluza--Klein theory: The strong version would suppose that the 
 $d$--dimensional inverse metric tensor is directly embedded
in the corresponding higher--dimensional inverse metric without
rescaling.
This would require a direct coupling of scalars
to the Einstein--Hilbert term in the low--dimensional theory.
The weak version would permit
field dependent reparametrizations between the actual $d$--dimensional
metric and the metric inherited from higher dimensions.

In the more recent history of physics, interest in the dilaton revived in the realm
of Kaluza--Klein supergravity. If the field content of space--time is 
assumed to arise from embeddings
in a $(4+d_i)$--dimensional manifold of Minkowski
signature with a compact factor, 
the $d_i$--dimensional internal manifold will have variable
volume from the four--dimensional point of view, and this variable volume will
be encoded in a local field $\phi(x)$, the dilaton.
However, my expectation is that a low--dimensional
Kaluza--Klein dilaton tells us only about
local variations of the volume of the internal manifold, its four--dimensional
dynamics can not fix the mean value of that volume. Explaining and fixing the 
smallness of  internal dimensions belongs to the realm of Kaluza--Klein 
cosmology or string theory
in the $(4+d_i)$--dimensional framework. Nevertheless, the
low--dimensional field theory must be self--consistent and
tell us why variations
of the internal dimensions are small or invisible, 
and this problem concerns the issue of
the effective potential of Kaluza--Klein type dilatons in four dimensions.

While Kaluza--Klein theory is not a suitable framework
for a unified theory of fundamental interactions, it still provides an
indispensable tool in field theoretical investigations of string theory, where
compact manifolds with or without boundaries arise both in low energy
effective field theories and in the net of extended objects with $p$ spatial
dimensions, so called $p$--branes.
The dynamics of nets of various $p$--branes
in $D\leq 11$ dimensions is currently under close investigation as a paradigm for 
non--perturbative effects in string theory. Far away
from the intersections the excitations of the $p$--branes
can be described in terms of $p+1$--dimensional
field theories on the branes, and compactness arises for those dimensions
which describe the extension of a given $p$--brane 
between other extended objects.

Quantum field theory on manifolds with compact directions
is an old subject of theoretical physics, with an almost canonical structure
in the bosonic sector and some matters of taste in the more complicated fermionic sector.
I will give an account on the compactification of one dimension in my 
favorite conventions, and the cogniscenti may find it amusing (or bothersome) to compare to 
their favorite frameworks.
We will go from $D$ to $D-1$ dimensions
through compactification on a circle, and assume $D\ge 4$ in the sequel.
As long as only zero modes of internal derivatives are taken into
account
repeated application of the following formulas can be used
to parametrize
any Kaluza--Klein theory ending up in $d\geq 3$ dimensions. However,
besides
compactifications from 5 to 4 and from 4 to 3 dimensions we will need
only a few selected results for compactifications from $D\geq 6$ to 4 dimensions.

In addition to serving as a generator for torus compactifications there
are further reasons to go from $D$ to $D-1$ dimensions:
My first motivation for this arose from 
recent developments in string theory: Motivated by the
succesful applications of duality symmetries in the investigation
of low energy effective actions of supersymmetric gauge theories, Witten observed
that theories with Bogomol'nyi saturated solitons should have dual
descriptions in terms of Kaluza--Klein theories, in order to explain the
equidistant mass spectrum of particles in the dual theory \cite{Wi1,Wi2}.
This fits very well with proposals by Witten
and by Ho\v{r}ava and Witten, saying that large dilaton
limits of type {\sl IIA} and heterotic $E_8\times E_8$
superstring theory are described by large radius
compactifications of an eleven--dimensional theory
on $M_{10}\times S^1$ or $M_{10}\times S^1/\mathbb{Z}_2$, which has been
called
M--theory \cite{Wi1,HW}.
While this initiated the current activity to define or identify M--theory
as a matrix theory or as a theory treating extended objects
of various dimensions as (almost)
equally fundamental degrees of freedom\footnote{My impression is that strings still
play a more fundamental role than other extended objects,
since their spectra determine
the brane--scan.}, 
it also motivated 
speculations about relations between supersymmetric
theories in three dimensions and non--supersymmetric theories 
in four dimensions, including in particular a proposal
to solve the cosmological constant
problem through supersymmetry in three dimensions \cite{Wi3}. 
While there is no complete picture yet concerning the impact of different
sectors of the moduli space of M--theory and string theory
on low energy physics, it implies 
existence of parametrizations of low energy effective field theory
which
implement a dilaton either in three or
in four dimensions\footnote{Even in the Einstein frame 
a radially coupled dilaton in a field theory
on $M_d\times X\times S^1$ 
looks like a radially coupled dilaton from an Einstein frame on $M_d\times S^1$
only in a very particular class of parametrizations of the Kaluza--Klein Ansatz.
Generically, the large dilaton limit will look like a scalar--vector--tensor theory
of gravity on $M_d\times S^1$.}.
It has been stressed by Banks and Dine 
that four--dimensional compactifications of
M--theory should imply a five--dimensional 
threshold two orders of magnitude below the GUT scale, 
i.e.\ around $10^{14}$ GeV \cite{BD1}, implying that quantum gravity
effects may become strong before unification is achieved in 
a field theoretic framework. In a recent update
on Kaplunovsky's work on large volume string compactifications
Caceres et al.\ also discuss a five--dimensional threshold 
in heterotic string theory \cite{CKM}.

A Kaluza--Klein decomposition of the metric which preserves the Einstein--frame is
\begin{equation}\label{KK}
G_{MN}=\Phi^{-\frac{1}{D-2}}\left(\begin{array}{cc} g_{\mu\nu}+\Phi a_\mu a_\nu
& \,\Phi a_\mu \\ \Phi a_\nu & \,\Phi\end{array}\right).
\end{equation}
A priori this is a mere reparametrization of the metric comparable to an ADM 
decomposition.
It becomes an Ansatz if we assume that the fields do not depend on $x^{D-1}$
or their dependence on $x^{D-1}$ is negligible, due to translational invariance
or compactness along $x^{D-1}$. Eq.\ (\ref{KK}) then
yields for the zero modes of
$\partial_{D-1}$ and up to surface terms\footnote{Whenever equations are 
written down
for terms appearing in a Lagrangian, we will consider expressions equivalent if they
differ by a divergence and neglect the divergences in the sequel. 

Compactification to the Einstein frame becomes singular for $D=3$, and
other Ans\"atze have to be employed in this case, see e.g.\
 \cite{BP,hn,NKS}.}
\begin{equation}
\sqrt{-G}\,G^{MN}R_{MN}=\sqrt{-g}\,[g^{\mu\nu}R_{\mu\nu}-
\frac{D-3}{4(D-2)}g^{\mu\nu}\partial_\mu\ln\Phi\cdot
\partial_\nu\ln\Phi-\frac{\Phi}{4}f_{\mu\nu}f^{\mu\nu}],
\end{equation}
while reduction of the 
Yang--Mills field $A_M{}^k\to A_\mu{}^k$, $A^k$
yields
\begin{equation}
-\frac{1}{4}\sqrt{-G}\,F_{MN}{}^k F^{MN}{}_k=-\frac{1}{4}\sqrt{-g}\,
\Phi^{\frac{1}{D-2}}[B_{\mu\nu}{}^k B^{\mu\nu}{}_k
+\frac{2}{\Phi}g^{\mu\nu}
D_\mu A^k \cdot D_\nu A_k],
\end{equation}
\[
B_{\mu\nu}=F_{\mu\nu}+a_\mu D_\nu A-a_\nu D_\mu A.
\]

We follow the usual terminology to denote the mechanism
leading from Einstein gravity in $D$ dimensions to Einstein--Yang--Mills
theory in $D-1$ dimensions as compactification on a circle. 
This has to be qualified in three directions:
First one should keep in mind that Einstein gravity actually induces 
a scalar--vector--tensor type theory of gravity on submanifolds,
which is not strictly a scalar--vector--tensor theory since
the scalar and vector degrees of freedom also couple
directly to matter.
In the scalar--tensor sector the induced theory
would resemble a Jordan--Brans--Dicke theory 
as a consequence of
the fact that the metric on the submanifold induced from the metric $G$
is 
\[
\tilde{g}_{\mu\nu}=\Phi^{-\frac{1}{D-2}}(g_{\mu\nu}+\Phi a_\mu a_\nu).
\]
If we insist that Kaluza--Klein theory yields Einstein gravity in lower
dimensions we suppose that the physical metric on the submanifold
is not the metric inherited from the embedding space.
Stated differently, the geodesics traced out by test particles are not the
geodesics of the embedded submanifold.

The second remark concerns the naive picture of a product 
manifold ${\cal M}\times\cal X$,
with $\cal X$ compact. While this is a particular possibility considered by 
Kaluza--Klein theory,
it is not the most general setting. Generically, the compact manifold $\cal X$ 
acts as a fiber
in a total space which projects to $\cal M$. Indeed, in the case of a product manifold
the vector fields $a$ could be safely neglected if the Riemannian structure respects
the product structure: The connection coefficients $\Gamma^{D-1}{}_{\mu\nu}$
and  $\Gamma^{\mu}{}_{D-1\nu}$ are gauge equivalent to zero for arbitrary $\Phi$
if and only if $a$ is a gradient. This is equivalent to the requirement that vectors
tangent and normal to the internal dimensions are mapped to tangent and normal
vectors under parallel translation.

The third remark also concerns the graviphotons $a$: To lowest order
in scalar contributions to the metric these vector fields contribute
Maxwell terms to the low--dimensional Lagrangian. However, fermions
will be neutral with respect to these gauge fields. Instead, the graviphoton
mixes with gauge fields inherited from the higher--dimensional theory in such 
a way that the low--dimensional gauge field becomes neutral under 
diffeomorphisms normal to the submanifold. This can be seen explicitly
in the reduction of fermion contributions to the Lagrangian carried
out below.

A detailed discussion of the reduction of fermion terms 
requires a distincton between
even and odd values of $D$.
My favorite choice for embeddings and reductions of $\gamma$--matrices
is based on Weyl bases in even dimensions and Dirac bases in odd
dimensions. General discussions of properties of spinors
in arbitrary dimensions can be found in \cite{vNW1,cw1}.

If $D$ is {\bf even} and $\gamma_\mu$
is a basis of Dirac matrices in $D-1$ dimensions, a Weyl basis of Dirac
matrices in $D$ dimensions is given by
\[
\Gamma_0 = \left(\begin{array}{cc} 0 & \,1\\ 1 & \,0\end{array}\right)\]
\[
\Gamma_j =
\left(\begin{array}{cc} 0&-\gamma_0\gamma_j\\
\gamma_0\gamma_j&0\end{array}\right)\]
\[
\Gamma_{D-1} = \left(\begin{array}{cc} 0 & -\gamma_0\\ \gamma_0 &
0\end{array}\right).
\]
The corresponding analog of $\gamma_5$ is
\[
\Gamma_{D+1}=i^{\frac{D}{2}+1}\Gamma_0\cdot\Gamma_1\ldots\Gamma_{D-1}.
\]
If we start with a $(1+0)$--dimensional $\gamma$--matrix $\gamma_0=-1$, 
and with the conventions for $\gamma$--matrices in odd dimensions 
described below, $\Gamma_{D+1}$
takes the form
\[
\Gamma_{D+1} = \left(\begin{array}{cr} 1 & \, 0 \\ 0 & \, -1\end{array}\right).
\]

This embedding of $\gamma$--matrices 
provides a direct mapping between the spinor representation
in $D$ dimensions and both inequivalent representations in $D-1$
dimensions, and the appropriate Ansatz for the dimensional
reduction of the spinor field is
\[
\Psi=\Phi^{\frac{1}{4(D-2)}}\left(\begin{array}{c}\psi_+\\ \psi_-
\end{array}\right).
\]
In
a gauge $E^\mu{}_{D-1}=0$ for the $D$--{\sl bein} 
the dimensionally reduced action
reads
\begin{equation}
\sqrt{-G}\,\overline{\Psi}[E^M{}_A\Gamma^A(i\partial_M
+i\Omega_M +qA_M)-M]\Psi=
\end{equation}
\[
=
\sqrt{-g}\,[\overline{\psi}_+ e^\mu{}_a\gamma^a_+
(i\partial_\mu +i\omega_{+\mu}+qV_\mu)\psi_+
+
\overline{\psi}_- e^\mu{}_a\gamma^a_-(i\partial_\mu
+i\omega_{-\mu} +qV_\mu)\psi_-]
\]
\[
+q\sqrt{-g}\,\Phi^{-\frac{1}{2}}(\overline{\psi}_+ A\psi_+ +\overline{\psi}_-
A\psi_-)+M\sqrt{-g}\,\Phi^{-\frac{1}{2(D-2)}}(\psi_+^+\psi_- +\psi_-^+\psi_+)
\]
\[
-\frac{i}{8}\sqrt{-g}\,\Phi^{\frac{1}{2}}f_{ab}
(\overline{\psi}_+ \gamma_+^{ab}\psi_+ +\overline{\psi}_-
\gamma_-^{ab}\psi_-)
\]
with
$\gamma^0_\pm = \pm \gamma^0$, $\gamma^j_\pm =\gamma^j$, while
\[
\Omega_\mu =-\frac{1}{4}\Gamma^A\Gamma^B\Omega_{AB\mu}
\]  
and
\[
\omega_{\pm\mu} =-\frac{1}{4}\gamma^a_\pm\gamma^b_\pm\omega_{ab\mu}
\] 
denote the canonical spin connections
for the metrics $G$ and $g$, respectively.
The vector field $V_\mu$ appearing as a gauge potential in $D-1$ dimensions is
\[
V_\mu =A_\mu-a_\mu A, 
\]
\[
V_{\mu\nu}=B_{\mu\nu}-Af_{\mu\nu}.
\]
Note that $D_\mu A_j=\partial_\mu A_j-qA_\mu{}^i f_{ij}{}^k
A_k=\partial_\mu A_j-qV_\mu{}^i f_{ij}{}^k
A_k$.

If $D$ is {\bf odd} and $\gamma_\mu$
is a basis of Dirac matrices in $D-1$ dimensions, a basis of Dirac
matrices in $D$ dimensions is given by 
\[
\Gamma_0=\left(\begin{array}{cc} -1 & \,0 \\ 0 &\,1\end{array}\right) 
\]
\[
\Gamma_j=\gamma_j, \qquad 1\leq j\leq D-2
\]
\[
\Gamma_{D-1}=-i\gamma_0.
\]
We have
\[
i^{\frac{D+1}{2}+1}\Gamma_0\cdot\Gamma_1\ldots\Gamma_{D-1}=1
\]
and a second inequivalent basis is given 
by $-\Gamma_0$, $\Gamma_J$, $1\leq J\leq D-1$.

Contrary to the embedding for even $D$ described above, this embedding
does not provide a direct mapping between spinor representations of
the Lorentz group in $D$ and $D-1$ dimensions, and we have to compensate for 
that through an extra factor $\cal X$ in the Kaluza--Klein Ansatz for spinors:
\[
\Psi=\Phi^{\frac{1}{4(D-2)}} {\cal X}\cdot\psi,
\]
\[
{\cal X}^+\cdot\Gamma_0\cdot{\cal X}=\gamma_0,
\]
\[
{\cal X}^+\cdot\Gamma_j\cdot{\cal X}=\Gamma_j, \qquad 1\leq j\leq D-2.
\]
In the bases of $\gamma$--matrices employed here, $\cal X$ is realized explicitly as
\[
{\cal X}=\frac{1}{\sqrt{2}}\left(\begin{array}{cr} 1&\,-1\\ 1&\, 1\end{array}\right).
\]

Employing again a gauge $E^\mu{}_{D-1}=0$ for the $D$--{\sl bein} 
the reduced action reads
\begin{equation}
\sqrt{-G}\,\overline{\Psi}[E^M{}_A\Gamma^A(i\partial_M
+i\Omega_M +qA_M)-M]\Psi=
\end{equation}
\[
=
\sqrt{-g}\,[\overline{\psi} e^\mu{}_a\gamma^a
(i\partial_\mu +i\omega_{\mu}+qV_\mu)\psi]
-iq\sqrt{-g}\,\Phi^{-\frac{1}{2}}\overline{\psi}\gamma_D A\psi
\]
\[
-M\sqrt{-g}\,\Phi^{-\frac{1}{2(D-2)}}\overline{\psi}\psi
-\frac{1}{8}\sqrt{-g}\,\Phi^{\frac{1}{2}}f_{ab}
\overline{\psi}\gamma^{ab}\gamma_D\psi
\]
with $\Omega$ and $\omega$ denoting the canonical spin connections
for the metrics $G$ and $g$, respectively,
and again $ V_\mu =A_\mu-a_\mu A $.

Since time reversal in odd dimensions mixes the
two equivalence class of representations of the 
corresponding Clifford algebra,
one has to add a second spinor $\Psi_-$ related to $\Gamma_-^0=
-\Gamma^0$, $\Gamma_-^J=\Gamma^J$ and a corresponding 
low--dimensional spinor $\psi_-$ through
\[
\Psi_-=\Phi^{\frac{1}{4(D-2)}} {\cal X}_-\cdot\psi_-
\]
with
\[
{\cal X}_-=\frac{1}{\sqrt{2}}\left(\begin{array}{cc} 1&\,1\\ -1&\, 1\end{array}\right).
\]

This yields then the same low--dimensional action with the sign of the
parity violating terms inverted. To end up with an irreducible representation
of the Lorentz group in $D-1$ dimensions we impose the 
constraint $\psi_-=\psi$, whence
our $(D-1)$--dimensional action is
\[
{\cal L}=
\sqrt{-g}\,[\overline{\psi} e^\mu{}_a\gamma^a
(i\partial_\mu +i\omega_{\mu}+qV_\mu)\psi]
-M\sqrt{-g}\,\Phi^{-\frac{1}{2(D-2)}}\overline{\psi}\psi.
\]

Both for even and odd values of $D$ we have
not encountered a photon--like coupling of the
graviphoton to 
low--dimensional fermions. This is can be understood
very easily: The only place where a photon--type
coupling of the graviphoton
could arise on zero modes of $\partial_{D-1}$
is in the reduction of the spin--connection,
through terms containing 
\[
E^{D-1}{}_a=-\Phi^{\frac{1}{2(D-2)}}a_\mu e^\mu{}_a.
\]
However, all these terms contain derivatives 
on $\Phi$, $e_\mu{}^a$ or $a_\mu$ and can not create
a $U(1)$ gauge coupling. On the zero modes
components of the metric tensor in higher dimensions
modify gauge fields upon compactification, but they do
not create new gauge couplings to fermions beyond
those couplings already present in higher 
dimensions. Of course, a
possible way to avoid this negative verdict on Kaluza--Klein
generated gauge fields relies on eigenspinors
of the internal Dirac operator $E^{D-1}{}_a\gamma^a\partial_{D-1}$,
but this will play no role in the sequel.

In concluding this section I would like to add a remark
on the string frame in four dimensions: String theory
in the string frame supposes that the dilaton couplings
in four dimensions look like Kaluza-Klein couplings
with the four--dimensional
metric directly induced from 
an embedding space, i.e.\
it is based on the strong rather than the weak version
of Kaluza--Klein theory. Neglecting the gauge fields
and denoting by $\Phi_{\cal BD}^2$ the determinant of the internal
metric, compactification from $D$ dimensions
to the Einstein frame and
a string--like frame in four dimensions would proceed via
\[
G_{MN}=\Phi_{\cal BD}^{-1}\left(\begin{array}{cc} g_{\mu\nu}
& \, 0 \\ 0 & \,\Phi_{\cal BD}^{\frac{D-2}{D-4}}h_{mn}\end{array}\right)
=\left(\begin{array}{cc} \tilde{g}_{\mu\nu}
& \, 0 \\ 0 & \,\Phi_{\cal BD}^{\frac{2}{D-4}}h_{mn}\end{array}\right),
\]
respectively, and yield
\begin{eqnarray}
\sqrt{-G}\,G^{MN}R_{MN}&=&\sqrt{-g}\,g^{\mu\nu}(R_{\mu\nu}-
\frac{D-2}{2(D-4)}\partial_\mu\ln\Phi_{\cal BD}\cdot
\partial_\nu\ln\Phi_{\cal BD})\\ &=&
\sqrt{-\tilde{g}}\,\tilde{g}^{\mu\nu}\Phi_{\cal BD}(\tilde{R}_{\mu\nu}+
\frac{D-5}{D-4}\partial_\mu\ln\Phi_{\cal BD}\cdot
\partial_\nu\ln\Phi_{\cal BD}). \nonumber
\end{eqnarray}
Therefore,
in strong Kaluza--Klein theory gravity would be described by a 
Brans--Dicke type theory
with a Brans--Dicke parameter
\[
\omega_{\cal BD}=-\frac{D-5}{D-4}.
\]

This is not yet gravity in the string frame, since both $\Phi_{\cal BD}$ 
and the string dilaton $\phi_S$
couple to the Yang--Mills terms, and the metric $\tilde{g}_{\mu\nu}$ has 
to be Weyl rescaled 
by $\exp(\sqrt{\frac{\kappa}{2}}\phi_S)$
to ensure equal coupling  both to $R$ and to $F^2$. The resulting theory 
in the gravitational
sector is a Brans--Dicke theory (\ref{LBD}) with 
\[
\omega_{\cal BD}=-1,
\]
and initially $\Lambda(\Phi_{\cal BD})=0$. However, Brans--Dicke
theory is constrained by the fact
that solar system tests of gravity restrict Brans--Dicke parameters
for massless Brans--Dicke scalars to $\omega_{\cal BD}>500$ \cite{cmw}. 
This leaves two
possibilities for the string frame:  We either should identify a mechanism 
to generate 
a large mass for $\Phi_{\cal BD}$ if low energy gravity in string theory
is described by a scalar--tensor type theory in the string frame,
or higher loop effects in string theory modify the Brans--Dicke coupling function
from $R\Phi_{\cal BD}$ to a convex function $C(\Phi_{\cal BD})R$. In the second 
case the cosmological
attractor mechanism of Damour and Nordtvedt would apply \cite{DV}, 
and cosmological
evolution would restore effective Einstein gravity in the string frame.

Our main interest
in the present work is in light dilatons, and therefore we rely on the conservative
assumption that low energy gravity is described by Einstein gravity.

Compactification of heterotic string theory in the Einstein frame 
shows that the four--dimen\-sio\-nal
dilaton $\phi$ coupling to gauge fields arises as a linear combination of the 
string dilaton $\phi_S$ already present in the ten--dimensional field theory limit
and
a Kaluza--Klein dilaton $\phi_K$ arising from the compactification 
to four dimensions \cite{Wilog}. If 
both $\phi_S$ and $\phi_K$ are normalized to have standard kinetic terms
in four dimensions the dilaton is dominated
by the Kaluza--Klein component with a mixing angle $\theta_\phi=-\frac{\pi}{6}$:
\[
\phi=\frac{1}{2}(\sqrt{3}\phi_K-\phi_S),
\]
and its decay constant is
\begin{equation}\label{fmpl}
f_\phi=\frac{m_{Pl}}{\sqrt{2}}.
\end{equation}

This relies on a parametrization for the string dilaton such that
strong gauge coupling in ten dimensions corresponds to strongly coupled
string theory and is based on the fact that the ten--dimensional
field theory limit of heterotic string theory
consists of $N=1$ supergravity 
coupled to $N=1$ supersymmetric gauge theory \cite{GHMR2,GS,GSW}.
This theory contains pieces derived from eleven--dimensional
supergravity, but the string dilaton couples stronger than
a Kaluza--Klein dilaton from eleven dimensions, and for this reason the
dilaton decay constant in four dimensions realizes a lower bound for
Kaluza--Klein decay constants arising through compactifications
from $D$ dimensions:
\[
f_{\phi(KK)}=\frac{m_{Pl}}{\sqrt{2}}\sqrt{\frac{D-2}{D-4}}>f_\phi.
\]

The alert reader may wonder why 
neither in $f_\phi$ nor in any pure Kaluza--Klein decay constant
any compactification scales show up.
This is due to the fact that internal volumes rescale higher--dimensional
gravitational constants to the four--dimensional constant $\kappa$,
and in normalizing dilatons to standard kinetic terms only a rescaling 
with $m_{Pl}$ is involved. Therefore the decay constants only depend on
the Planck mass. 

A dilaton coupling scale (\ref{fmpl})
of the order of the Planck mass 
implies invisibility of the dilaton from the
particle physics point of view: 
We can readily calculate the integrated tree level cross section
for creation of a dilaton pair
through head--on collision of two gauge bosons with Mandelstam
parameter $s$
\[
\sigma=\frac{s}{64\pi f_\phi^4},
\]
and this tells us that even for Planck scale collisions the cross section
would be tiny $\sigma\simeq \frac{1}{4\pi}l_{Pl}^2$.

\newpage
\section{The dilaton in three dimensions}\label{dil3d}
 
Interest in a three--dimensional dilaton
arose from Witten's observation that theories with 
Bogomol'nyi saturated
solitons may be related to theories in 
higher dimensions in a Kaluza--Klein
type framework \cite{Wi1,Wi2}. 
This idea can be motivated from classical 
duality considerations
which generically imply a trading between 
solitons and particles
in dual theories. Given the soliton--particle 
correspondence and the infinite
tower of equidistant solitonic excitations 
it seems very natural to 
relate
solitons with an equidistant mass spectrum
to compactified
theories
in higher dimensions.
In this framework the particular case of 
dualities between
supersymmetric theories in 2+1 dimensions and 
non--supersymmetric
theories in 3+1 dimensions deserves special 
attention, since the
four--dimensional theory
might inherit the vanishing of the cosmological 
constant from the 
corresponding
three--dimensional theory\footnote{Supersymmetry as 
a solution to the
cosmological constant problem has been discussed
in \cite{BZ}. A very useful review and critical 
discussion of several
attempts to solve the problem can be found in \cite{sw2}.}
\cite{Wi2,Wi3}.
A discussion of the supersymmetric abelian 
Higgs model in 2+1 
dimensions
coupled to supergravity confirmed this 
picture by showing that the soliton
spectrum in this theory is not 
supersymmetric \cite{BBS}.
It may also be worth--while to point out 
that due to the topological nature of 
gravity in three dimensions
one would not need a fully fledged
supergravity multiplet to get rid of the 
cosmological constant in this
scenario.
In particular, we would not need a gravitino, 
which would be hard to 
accomodate in the four--dimensional theory.

However, soon after Witten's proposal
worries arose that the
static potential
in 2+1 dimensions would imply a logarithmic 
divergence of the dilaton
for any static source. 
In spite of that it turned out that the 
infrared singularity 
of the propagator actually suppresses fluctuations
of the dilaton in three dimensions.
The suppression of fluctuations works because
fermions and adjoint scalars
provide sources for the dilaton which differ 
in sign from the
dilaton sources provided by the gauge fields 
arising from
the four--dimensional metric and four--dimensional 
gluons. Finite
energy or independence of the theory from an infrared
regulator
then implies that any local dilaton source 
has to be compensated
by another dilaton source somewhere else, whence the 
dilaton vanishes
asymptotically and the radius of the internal dimension 
would approach a value to be
fixed by string theory. 

The idea to get rid of a dilaton through duality symmetries
between theories in different dimensions is very speculative,
but the mechanism outlined here \cite{rdplb1} provides an
example for non--perturbative stabilization of a dilaton
in a low energy
effective theory different from the mechanisms
outlined in section \ref{stab4d}.

In order to make the observations outlined above quantitative, 
I will 
take a four--dimensional point of view and
discuss the
action of the three--dimensional
dilaton arising through a Kaluza--Klein parametrization
of Einstein--Yang--Mills theory in four dimensions.

We relate the dilaton $\phi$ to the metric coefficient $\Phi$
via
\[
\Phi=\exp(\sqrt{8\kappa}\phi),
\]
where the three--dimensional gravitational constant $\kappa$
is the four--dimensional gravitational constant divided by
the circumference of the compact dimension.
After appropriate rescalings of the other
fields and coupling constants,
we infer the following action in three dimensions from the results
of the previous section:
\[
\frac{1}{\sqrt{-g}}{\cal L}=\frac{1}{2\kappa}R-\frac{1}{2}g^{\mu\nu}
\partial_\mu\phi\cdot\partial_\nu\phi-\frac{1}{4}\exp(\sqrt{8\kappa}\phi)
f_{\mu\nu}f^{\mu\nu}
\]
\[
-\frac{1}{4}\exp(\sqrt{2\kappa}\phi)
(V_{\mu\nu}{}^k V^{\mu\nu}{}_k +2\sqrt{2\kappa}V_{\mu\nu}{}^k
A_k f^{\mu\nu}+2\kappa A_k A^k f_{\mu\nu}f^{\mu\nu})
\]
\[
+\overline{\psi}_+ e^\mu{}_a\gamma^a_+
(i\partial_\mu +i\omega_{+\mu}+qV_\mu)\psi_+
+
\overline{\psi}_- e^\mu{}_a\gamma^a_-(i\partial_\mu
+i\omega_{-\mu} +qV_\mu)\psi_-
\]
\[
+q\exp(-\sqrt{2\kappa}\phi)
(\overline{\psi}_+ A\psi_+ +\overline{\psi}_-
A\psi_-)+M\exp\!\Big(-\sqrt{\frac{\kappa}{2}}\phi\Big)(\psi_+^+\psi_- +\psi_-^+\psi_+)
\]
\[
-\frac{1}{2}\exp(-\sqrt{2\kappa}\phi)g^{\mu\nu}D_\mu A_k
D_\nu A^k
-\frac{i}{8}\sqrt{2\kappa}\exp(\sqrt{2\kappa}\phi)f_{ab}
(\overline{\psi}_+ \gamma_+^{ab}\psi_+ +\overline{\psi}_-
\gamma_-^{ab}\psi_-).
\]

In order to evaluate the effect of the infrared divergence of the
electrostatic potential in 2+1 dimensions, we consider the energy
of a static configuration with the fermions in stationary orbits,
and in gauge $A_0=a_0=0$ (complying with $E^\mu{}_3=0$, since we 
employ
diffeomorphisms  which are constant along the normal direction):
\[
\frac{1}{\sqrt{-g}}{\cal H}=\frac{1}{2}g^{ij}
\partial_i\phi\cdot\partial_j\phi
+\frac{1}{4}\exp(\sqrt{2\kappa}\phi)
(V_{ij}{}^k  +\sqrt{2\kappa}A^k f_{ij})
(V^{ij}{}_k+\sqrt{2\kappa} A_k f^{ij})
\]
\[
+\frac{1}{4}\exp(\sqrt{8\kappa}\phi)
f_{ij}f^{ij}
-\overline{\psi}_+ e^j{}_a\gamma^a_+
(i\partial_j +i\omega_{+j}+qV_j)\psi_+
-
\overline{\psi}_- e^j{}_a\gamma^a_-(i\partial_j
+i\omega_{-j} +qV_j)\psi_-
\]
\[
-q\exp(-\sqrt{2\kappa}\phi)
(\overline{\psi}_+ A\psi_+ +\overline{\psi}_-
A\psi_-)-M\exp\!\Big(-\sqrt{\frac{\kappa}{2}}\phi\Big)(\psi_+^+\psi_- +\psi_-^+\psi_+)
\]
\[
+\frac{1}{2}\exp(-\sqrt{2\kappa}\phi)g^{ij}D_i A_k
D_j A^k
+\frac{i}{8}\sqrt{2\kappa}\exp(\sqrt{2\kappa}\phi)f_{ab}
(\overline{\psi}_+ \gamma_+^{ab}\psi_+ +\overline{\psi}_-
\gamma_-^{ab}\psi_-),
\]
which tells us that the graviphoton and the 
Yang--Mills fields
yield positive sources for the dilaton, while the kinetic
term of $A$ provides a negative contribution.
There is some ambiguity with regard to the contribution 
due to the fermions.
However, on--shell the fermion contribution adds up to a 
positive
term. Therefore, generically the fermions will contribute 
negative sources
to the dilaton.

This appearance of positive and negative source terms for 
the dilaton
apparently works for any value of $D$.
In ten--dimensional low energy effective actions of
superstring theory involving only $N=1$ supergravity
the dilaton couples only with one sign 
since eleven--dimensional
supergravity does not contain elementary fermions
and Yang--Mills fields, and the residual 3--form potential 
and graviphoton are
set to zero.

However, the case $D=4$ is peculiar due to the logarithmic 
IR divergence of the
electrostatic potential in 2+1 dimensions. In a linear 
approximation
the dilaton $\ln\Phi$ behaves like a massless scalar field
coupled to external sources,
whence finiteness of energy requires a vanishing dilaton charge:
\begin{equation}\label{cond}
\int d^2 \mbox{\bf x}\,\varrho(\mbox{\bf x})=
\int d^2 \mbox{\bf x}\,\Big(\sqrt{\frac{\kappa}{2}}f_{ij}f^{ij}
+\sqrt{\frac{\kappa}{8}}
(V+\sqrt{2\kappa}Af)_{ij}{}^k (V+\sqrt{2\kappa}Af)^{ij}{}_k
\end{equation}
\[
-\sqrt{\frac{\kappa}{2}}
g^{ij}D_i A^k \cdot D_j A_k
+\sqrt{2\kappa}q(\overline{\psi}_+ A\psi_+ +\overline{\psi}_-
A\psi_-)
\]
\[
+\sqrt{\frac{\kappa}{2}}M(\psi_+^+\psi_- +\psi_-^+\psi_+)
+\frac{i}{4}\kappa f_{ab}
(\overline{\psi}_+ \gamma_+^{ab}\psi_+ +\overline{\psi}_-
\gamma_-^{ab}\psi_-)\Big)
=0.
\]

This property is similar to charge neutrality of the Coulomb gas in
two dimensions and can be derived from
conformal invariance of the
partition function or independence of the length scale $\lambda$ entering
the definition of the electrostatic potential: 
\[
\Phi(\mbox{\bf x})=
\exp\!\Bigg(\frac{\sqrt{2\kappa}}{\pi}\int d^2\mbox{\bf x}'
\varrho(\mbox{\bf x}')\ln\frac{|\mbox{\bf x}-\mbox{\bf x}'|}{\lambda}\Bigg).
\]

Eq.\ (\ref{cond}) is equivalent to absence of a logarithmic 
singularity
of the dilaton.

If one feels uncomfortable about the use of a linear approximation
in this argument one may alternatively rely on independence
of the perturbation theory on the scale $\lambda$.
 $\lambda$ appears as an infrared cutoff if the static result 
is inferred
from the retarded potential
\[
G(t-t',\mbox{\bf x}-\mbox{\bf x}')=
\frac{\Theta\left((t-t')-|\mbox{\bf x}-\mbox{\bf x}'|\right)}{2\pi
\sqrt{(t-t')^2-|\mbox{\bf x}-\mbox{\bf x}'|^2}}.
\]

The self--trapping mechanism encoded in (\ref{cond})
is superficially stable with respect to quantum effects,
since calculation of the 1--loop
effective potential in 
dimensional regularization yields no perturbatively
generated effective potential. This is explained for four dimensions
in appendix B.
I would also like to point out
that the physics behind (\ref{cond}) is more transparent than 
in the case
of the Coulomb gas, since particles and antiparticles contribute in
the same way to the dilaton: If a gauge boson excites a 
dilaton field
the divergence of the resulting energy density implies
pair production of adjoint scalars and fermions to restore 
an asymptotically
vanishing dilaton. Clearly, $M$ suppresses
the production of fermions relative to adjoint scalars.
Stated in another way: The adjoint scalar and light fermions screen
the dilaton charge of the gauge bosons.

Compactification to three dimensions
and subsequent decompactification due to BPS solitons
is an interesting,
but speculative proposal for the low energy sector of string theory,
and we will concentrate on the four--dimensional dilaton
in the sequel.

\newpage
\section{Generalized Coulomb potentials in gauge theory with a dilaton}\label{copot}

As pointed out in the previous sections we expect dilatonic degrees of freedom in
four--dimensional gauge theories if physics at very high energies
involves decompactification of internal dimensions or string theory.
To acquire a better understanding of the impact of dilatons in four--dimensional
gauge theory we now look into the problem how a light dilaton modifies
the Coulomb potential and its non--abelian analog \cite{rdplb2}. It turns out
that the dilaton introduces an ambiguity due to different boundary conditions
which can be imposed on the dilaton: Two interesting solutions which
arise include a regularized potential proportional to $(r+r_\phi)^{-1}$, where
 $r_\phi$ is inverse proportional to the decay constant of the dilaton, and
a confining potential proportional to $r$.

Here we are interested in low energy gauge theories, i.e.\ in the dynamics
of initially massless modes from the point of view of string theory.
Since the compactification scale or string scale are many orders
of magnitude larger than the weak scale, where the low energy
degrees of freedom described in the standard  model of particle physics
acquire their masses,
we do not expect the dilaton to couple to the relevant masses
at the weak scale. 
Modulo an effective potential which the dilaton may have acquired
on the road down from the string/compactification scales to temperatures
below the SUSY scale, the influence of a dilaton on  a low energy gauge theory
is then described by a Lagrange density
\begin{equation}\label{lagden}
{\cal L}=-\frac{1}{4}\exp(\frac{\phi}{f_\phi})F_{\mu\nu}{}^j F^{\mu\nu}{}_j
-\frac{1}{2}\partial^\mu\phi\cdot\partial_\mu\phi+\sum_{f=1}^{N_f}\overline{\psi}_f
(i\gamma^\mu\partial_\mu +q\gamma^\mu A_\mu{}^j X_j -m_f)\psi_f,
\end{equation}
with $X_j$ denoting a defining $N_c$--dimensional representation
of su$(N_c)$. 

I already set the axion to zero, since the static pointlike source
considered below does not excite the axion field.

The equations of motion are
\begin{equation}\label{eqmot1} 
\partial_\mu\Big(\exp(\frac{\phi}{f_\phi})F^{\mu\nu}{}_i\Big)
+q\exp(\frac{\phi}{f_\phi})A_\mu{}^j f_{ij}{}^k F^{\mu\nu}{}_k=
-q\overline{\psi}\gamma^\nu X_i\psi,
\end{equation}
\begin{equation}\label{eqmot2}
\partial^2\phi=\frac{1}{4f_\phi}\exp(\frac{\phi}{f_\phi})F_{\mu\nu}{}^j F^{\mu\nu}{}_j,
\end{equation}
\begin{equation}\label{eqmot3}
(i\gamma^\mu\partial_\mu +q\gamma^\mu A_\mu{}^j X_j-m)\psi=0,
\end{equation}
where here and in the sequel flavor indices are suppressed.

To analyze eq.\ (\ref{eqmot1}) we will find it convenient to rewrite it
in terms of the chromo--electric and magnetic fields $E_i=-F_{0i}{}^j X_j$,
 $B^i=\frac{1}{2}\epsilon^{ijk}F_{jk}{}^l X_l$:
\[
\nabla\cdot\Big(\exp(\frac{\phi}{f_\phi})\vek{E}\Big)
-iq\exp(\frac{\phi}{f_\phi})(\vek{A}\cdot\vek{E}
-\vek{E}\cdot\vek{A})=\varrho,
\]
\[
\partial_0\Big(\exp(\frac{\phi}{f_\phi})\vek{E}\Big)
-\nabla\times\Big(\exp(\frac{\phi}{f_\phi})\vek{B}\Big)
+iq\exp(\frac{\phi}{f_\phi})([\Phi,\vek{E}]+\vek{A}\times\vek{B}
+\vek{B}\times\vek{A})=-\vek{j}
\]
\[
\partial_0\vek{B}+iq[\Phi,\vek{B}]+
\nabla\times\vek{E}-iq(\vek{A}\times\vek{E}
+\vek{E}\times\vek{A})=0,
\]
\[
\nabla\cdot\vek{B}-iq(\vek{A}\cdot\vek{B}
-\vek{B}\cdot\vek{A})=0,
\]
where in the gauge theory 
above $\varrho=q(\psi^+\cdot X_k\cdot \psi)X^k$, 
 $j_i=q(\overline{\psi}\cdot\gamma_i X_k\cdot\psi)X^k$,
and we have included the Bianchi identities. In this section  
we use the letter $\Phi$ for $A^0$.

To discuss the impact of the dilaton on the Coulomb potential 
we consider static configurations: $\partial_0\varrho=0$, $\vek{j}=0$.
Then we learn from $\partial_\mu j^\mu-iq[A_\mu,j^\mu]=0$ that $\Phi$ 
and $\varrho$
are in the same Cartan subalgebra: $[\Phi,\varrho]=0$.

Pointlike stationary charge distributions, which in the present setting 
give rise to the generalized
Coulomb potentials, are special cases of SU($N_c$) currents of the form
\begin{equation}\label{rhosep}
j^\mu(x)=\varrho^i(\vek{r})X_i\eta^\mu{}_0=\rho(\vek{r})C^i X_i\eta^\mu{}_0
\end{equation}
carrying the same $\vek{r}$--dependence along any direction in color space.
Such distributions arise for separable quark 
wave functions $\psi(x)=\varphi(x)\zeta$, where $\zeta$
is a constant Lorentz scalar in a spinor 
representation of SU($N_c$), and $\varphi(x)$ is 
a SU($N_c$)--invariant
Dirac spinor whose  
left and right handed components differ only by a phase.
We also assume both factors normalized 
according to $\int d^3\vek{r}\,\overline{\varphi}\cdot\varphi=1$,
 $\zeta^+\cdot\zeta=1$.

For SU($N_c$) charges of the form (\ref{rhosep}) the vector potential can consistently
be neglected, whence $\vek{E}=-\nabla\Phi$ and the Yang--Mills equations reduce to
\[
\nabla\cdot\Big(\exp(\frac{\phi}{f_\phi})\nabla\Phi\Big)=-\varrho,
\]
\[
[\Phi,\nabla\Phi]=0.
\]
Due to (\ref{rhosep}) the second equation is fulfilled as a consequence of the first
equation.

Our aim is to determine the chromo--electric potential 
for a point charge
\[
\varrho_i(\vek{r})=qC_i\delta(\vek{r})
\]
where $C_i$ denotes the expectation value of the generator $X_i$
in color space. From the relation
\begin{equation}\label{su3}
(X_i)_{ab}(X^i)_{cd}=
\frac{1}{2}\delta_{ad}\delta_{bc}-\frac{1}{2N_c}\delta_{ab}\delta_{cd}
\end{equation}
one finds for arbitrary color content
\[
\sum_{i=1}^{N_c^2-1} C_i^2=\frac{N_c-1}{2N_c}.
\]

We thus want to determine the field of a stationary pointlike quark from
\begin{equation}\label{gl2}
\nabla\cdot\Big(\exp(\frac{\phi(\vek{r})}{f_\phi})\vek{E}_i(\vek{r})\Big)
=qC_i\delta(\vek{r}),
\end{equation}
\begin{equation}\label{fl2}
\nabla\times\vek{E}_i(\vek{r})=0,
\end{equation}
and
\begin{equation}\label{dil}
\Delta\phi(\vek{r})=-\frac{1}{2f_\phi}\exp(\frac{\phi(\vek{r})}{f_\phi})\vek{E}_i(\vek{r})
\cdot\vek{E}^i(\vek{r}).
\end{equation}

The unique radially symmetric solution to (\ref{gl2}) can be written down immediately:
\begin{equation}\label{sol1}
\exp(\frac{\phi(r)}{f_\phi})\vek{E}_i(\vek{r})
=\exp(\frac{\phi(r)}{f_\phi})E_i(r)\vek{e}_r=
\frac{qC_i}{4\pi r^2}\vek{e}_r
\end{equation}
whence equation (\ref{fl2}) is also satisfied. Equation (\ref{dil}) then translates into
\begin{equation}\label{dil2}
\frac{d^2}{dr^2}\phi(r)+\frac{2}{r}\frac{d}{dr}\phi(r)=
-\frac{q^2}{64\pi^2f_\phi}\Big(1-\frac{1}{N_c}\Big)
\exp\!\Big(-\frac{\phi(r)}{f_\phi}\Big)\frac{1}{r^4}.
\end{equation}

The form of this equation suggests an ansatz $\frac{\phi(r)}{f_\phi}=a\ln(\frac{r}{b})$,
which yields the solution discussed below.
However, we can solve (\ref{dil2}) for arbitrary boundary conditions 
through a substitution 
\begin{equation}\label{sub}
\xi=\frac{q}{4\pi f_\phi r}\sqrt{\frac{1}{2}-\frac{1}{2N_c}}, \qquad
\theta(\xi)=\frac{\phi(r)}{f_\phi},
\end{equation}
yielding\footnote{We can map the dilaton equation of motion for
arbitrary number $d$ of spatial dimensions to eq.\ (\ref{eqpsi})
through the substitution
\[
\xi=\frac{q}{f_\phi}\sqrt{\frac{1}{2}-\frac{1}{2N_c}}G_d(r)
\]
with
\[
G_d(r)=-\frac{1}{2\pi}\ln(\frac{r}{r_0}),\qquad d=2,
\]
\[
G_d(r)=\frac{\Gamma(\frac{d}{2})}{2(d-2)\sqrt{\pi}^d}\frac{1}{r^{d-2}},\qquad d>2.
\]}
\begin{equation}\label{eqpsi}
\frac{d^2}{d\xi^2}\theta(\xi)=-\frac{1}{2}\exp(-\theta(\xi)),
\end{equation}
or in terms of boundary conditions at infinity:
\begin{equation}\label{eqpsi2}
\theta'(\xi)^2-\theta'(0)^2=\exp(-\theta(\xi))-\exp(-\theta(0)),
\end{equation}
\[
\xi=
\int_{\theta(0)}^{\theta(\xi)}
\frac{d\theta}{\sqrt{\exp(-\theta)-\exp(-\theta(0))+\theta'(0)^2}},
\]
where a sign ambiguity has been resolved by the requirement that the dilaton
should not diverge at finite radius. The integral can be done elementary, with
two branches depending on the sign of $\theta'(0)^2-\exp(-\theta(0))$.

The presence of the dilaton introduced a two--fold ambiguity in the
Coulomb problem, and we have to determine from physical requirements
which boundary conditions to chose.

 For a first solution we require
that the dilaton generated by the pointlike quark vanishes at infinity while the
gradient satisfies the minimality condition
\begin{equation}\label{mincon}
\lim_{r\to \infty}r^2\frac{d}{dr}\phi(r)=
-\frac{q}{4\pi}\sqrt{\frac{1}{2}-\frac{1}{2N_c}}.
\end{equation}
This gives minimal kinetic energy for the dilaton at infinity subject to the constraint
that the 
chromo--electric field does not develop a singularity for positive finite $r$. 
Then we find for the radial dependence of the dilaton and the electric field
\begin{equation}\label{dilsol}
\phi(r)=2f_\phi
\ln\!\Big(1+\frac{q}{8\pi f_\phi r}\sqrt{\frac{1}{2}-\frac{1}{2N_c}}\Big),
\end{equation}
\begin{equation}\label{esol}
\vek{E}_i(\vek{r})=
\frac{qC_i}{4\pi\Big(r+\frac{q}{8\pi f_\phi}\sqrt{\frac{1}{2}
-\frac{1}{2N_c}}\Big)^2}\vek{e}_r,
\end{equation}
implying a modified Coulomb potential
\begin{equation}\label{coulpot}
\Phi_i(r)=\frac{qC_i}{4\pi r+\frac{q}{2 f_\phi}\sqrt{\frac{1}{2}-\frac{1}{2N_c}}}.
\end{equation}

The result for gauge group U(1) is received through the substitution $N_c\to -1$,
and the corresponding dilaton--photon configuration was proposed already
as a solitonic solution in a remarkable paper
by Cveti\v{c} and Tseytlin \cite{CT}.

The removal of the short distance singularity in the chromo--electric field would imply
finite energy of the dilaton--gluon configuration:
\begin{equation}\label{selferg}
E=\int d^3\vek{r}
\Big(\frac{1}{2}\nabla\phi\cdot\nabla\phi+\frac{1}{2}\exp(\frac{\phi(\vek{r})}{f_\phi})
\vek{E}_i(\vek{r})
\cdot\vek{E}^i(\vek{r})\Big)=2qf_\phi\sqrt{\frac{1}{2}-\frac{1}{2N_c}}.
\end{equation}

This regularization of the Coulomb potential at high energies
is a very attractive property: In view of the prediction
 (\ref{fmpl}) it means that the dilaton resolves pointlike singularities
at the Planck scale, fitting very well with the interpretation of the dilaton as a
low energy imprint of a theory of quantum gravity. 

It is clear that even a dilaton photon coupling and a resulting regularization
of the electromagnetic Coulomb potential at or below $l_{Pl}$ would not 
contradict accelerator based ``confirmations'' of pointlike structure of electrons
above $10^{-3}$ fm $\simeq 10^{16}\,l_{Pl}$, and
similarly the dilaton would also not show up spectroscopically:
Applied to the hydrogen atom, the regularization (\ref{coulpot}) would
imply a shift of energy levels of order
\[
\frac{\Delta E}{E}\simeq 
-\sqrt{\frac{\alpha_{em}}{8\pi}}\frac{l_{Pl}}{a_B}\simeq -2.6\times 10^{-28}.
\]
 For comparison, the 1S--2S level splitting and the Rydberg constant 
are known with a relative precision of order $10^{-11}$ \cite{AKWL}, and counted in terms
of orders of magnitude a dilaton regularized electromagnetic Coulomb potential
is as invisible
in high energy physics experiments as it is spectroscopically.

However,
besides (\ref{coulpot}) there exists another quite intriguing solution if we require
that $\frac{1}{q^2}\exp(\frac{\phi}{f_\phi})$ is independent of $q$.
This requirement arises naturally in string theory, since the non--perturbatively
fixed expectation value of the dilaton itself
is supposed to determine the coupling. In the
action (\ref{lagden}) this requirement amounts to the constraint that
the solution should respect the scale invariance of the equations
of motion under
\[
\phi\to\phi+2\eta f_\phi
\]
\[
A\to \exp(-\eta)A
\]
\[
q\to \exp(\eta)q
\] 
for constant $\eta$.
Eqs.\ (\ref{sub},\ref{eqpsi2}) then imply $\theta'(\xi)^2=\exp(-\theta(\xi))=4\xi^{-2}$, 
yielding
\begin{equation}\label{dilsol2}
\phi(r)=
2f_\phi\ln\!\Big(\frac{q}{8\pi f_\phi r}\sqrt{\frac{1}{2}-\frac{1}{2N_c}}\Big),
\end{equation}
\begin{equation}\label{esol2}
\vek{E}_i(\vek{r})=
\frac{32\pi f_\phi^2}{q}\frac{N_c}{N_c-1}C_i\vek{e}_r.
\end{equation}
This corresponds to an energy density 
\[
{\cal H}(\vek{r})=4\frac{f_\phi^2}{r^2}
\]
whence the energy in a volume of radius $r$ diverges linearly:
\[
E|_r=16\pi f_\phi^2 r.
\]
This is an infrared divergence, whence it should not be related
to new physics at short distances, and
it would cost an infinite amount of energy to create an isolated quark.
Gauge theory with a dilaton thus accomodates both Coulomb and confining
phases in a simple way.

\newpage
\section{The axidilaton and stabilization of the dilaton 
in four dimensions}\label{stab4d}

Yet we have been missing the pseudo--scalar axion which couples
to the instanton density $F\cdot\tilde{F}$. The motivation for including
an axion in theories with a dilaton is four--fold: Historically the first and still a
very important motivation for the axion arose from the observation that it
explains the absence of a CP violating phase in gauge theories \cite{PQ,sw1,fw2}.
Besides this an axion arises also as a massless excitation of closed superstrings
as a companion of the graviton and the dilaton \cite{GSW}, and it 
accompanies the dilaton
in supersymmetric gauge theories: If the dilaton arises in the real part
of the lowest component of a chiral superfield the corresponding imaginary
part is an axion. Furthermore, under a certain
constraint on decay constants the axion--dilaton system exhibits
a duality 
symmetry commonly denoted as S--duality. This has been realized both
in field theory \cite{STW} and in string theory\footnote{Target space
duality mixes dilaton-- and axion--like degrees of freedom in string theory
in a similar manner, see \cite{FLT,FILQ1} and references there.}
 \cite{FILQ2,jhs,ScSe,sen}. The 
dilaton and the
axion are mixed under this symmetry in a non--linear way, 
and recent developments in string theory indicate that S--duality
should be a generic feature of grand unified quantum field theories inherited from
string theory.

The primary motivations for contemporary 
considerations of a dilaton arise from string theory,
and in this spirit emphasis in the present 
discussion will also be on a string inspired axion.
The difference does not show up in the coupling to gauge fields, but in the vacuum
sector: A Peccei--Quinn type axion is an angular variable and has at most 
finitely many different vacua. 
The string axion on the other hand arises as a dual field to an antisymmetric
tensor and has no reason to be periodic. It also will not necessarily
couple to light
 fermion masses, yet it still suppresses a CP violating $\theta$--angle
in non--abelian gauge theory. To explain how this comes about, note that
an axion explains absence of CP violation with or without Peccei--Quinn
symmetry in the fermionic sector: If we temporarily include the axion scale
in the axion $\Theta(x)$, which thus becomes dimensionless,
the relevant axion--gluon term in the presence of $\theta$ is
\[
\frac{q^2}{32\pi^2}(\Theta+\theta)F^{\mu\nu}{}_j\tilde{F}_{\mu\nu}{}^j.
\]
However,
instantons always induce an effective axion potential such that
the vacuum expectation values of the axion 
satisfy 
 $\langle \Theta\rangle+\theta=2\pi n$ for some integer $n$, and this eliminates CP
violation from the $F\tilde{F}$--term. From this point of view Peccei--Quinn
symmetry on the fermions
only introduces an additional Higgs field, in order to derive the relevant
interaction indirectly through an anomaly. 

In the spirit of
employing Kaluza--Klein theory as a paradigm for theories with a dilaton
we first consider the axion
from a five--dimensional point of view:
In the presence of a five--dimensional
threshold the axion should arise as the fifth component of a
pseudo--vector. The Kaluza--Klein Ansatz (\ref{KK}) then yields
\[
\frac{q^2}{64\pi^2}\sqrt{-G}\epsilon^{JKLMN}\Theta_J F_{KL}{}^j
F_{MNj}=
\frac{q^2}{64\pi^2}\sqrt{-g}\epsilon^{\mu\nu\rho\sigma}[\Theta F_{\mu\nu}{}^j
F_{\rho\sigma j}+4\Theta_\mu F_{\nu\rho j}D_\sigma A^j],
\]
where $\epsilon_{01234}=-\sqrt{-G}$, i.e.\ $\epsilon$ is a tensor, not a density.
Since this term transforms into a divergence 
under $\Theta_K\to\Theta_K+\partial_K\alpha$, it is natural to propose a Maxwell term
as kinetic term:
\begin{equation}\label{maxax}
-\frac{1}{4}\sqrt{-G}\,\Theta_{MN}\Theta^{MN}=-\frac{1}{4}\sqrt{-g}\,
\Phi^{\frac{1}{3}}[b_{\mu\nu} b^{\mu\nu}
+\frac{2}{\Phi}g^{\mu\nu}
\partial_\mu \Theta \cdot \partial_\nu\Theta ],
\end{equation}
where $\Theta_{MN}=\partial_M\Theta_N-\partial_N\Theta_M$ and
\[
b_{\mu\nu}=\Theta_{\mu\nu}+\sqrt{2\kappa}a_\mu \partial_\nu\Theta-
\sqrt{2\kappa}a_\nu \partial_\mu\Theta.
\]
Here we have rescaled the graviphoton $a_\mu\to \sqrt{2\kappa}a_\mu$
to have canonical mass dimension.

It is an interesting property of a five--dimensional threshold that
the power of the dilaton in front of the kinetic term of the axion $\Theta$
is such that
it matches exactly with the SL(2,$\mathbb{R}$) duality
for the axidilaton system described below (\ref{dual}). This is a unique
property of reductions from five to
four dimensions and in remarkable coincidence with expectations from
string theory.

In the sequel we will use the following pseudo--vector
in four dimensions:
\[
\overline{\Theta}_\mu=\Theta_\mu-\sqrt{2\kappa}a_\mu\Theta,
\]
\[
\overline{\Theta}_{\mu\nu}=b_{\mu\nu}-\sqrt{2\kappa}\Theta f_{\mu\nu}.
\]

In order to motivate the following investigations, we take a closer look
at the action of the zero modes of five--dimensional
Einstein--Yang--Mills theory with fermions compactified to four dimensions.
We relate the dilaton $\phi$ to the metric coefficient $\Phi$
by
\[
\Phi=\exp(\sqrt{6\kappa}\phi),
\]
where the four--dimensional gravitational constant $\kappa$
is the five--dimensional gravitational constant divided by
the circumference of the compact dimension.
After appropriate rescalings of the other
fields and coupling constants,
we infer the following action in four dimensions from the results
of section \ref{kaka}:
\[
\frac{1}{\sqrt{-g}}{\cal L}=\frac{1}{2\kappa}R-\frac{1}{2}g^{\mu\nu}
\partial_\mu\phi\cdot\partial_\nu\phi-\frac{1}{4}\exp(\sqrt{6\kappa}\phi)
f_{\mu\nu}f^{\mu\nu}
\]
\[
-\frac{1}{2}\exp\!\Big(-\sqrt{\frac{8\kappa}{3}}\phi\Big)g^{\mu\nu}(D_\mu A_j\cdot
D_\nu A^j+\partial_\mu\Theta\cdot\partial_\nu\Theta)
\]
\[
-\frac{1}{4}\exp\!\Big(\sqrt{\frac{2\kappa}{3}}\phi\Big)
[V_{\mu\nu}{}^j V^{\mu\nu}{}_j 
+\overline{\Theta}_{\mu\nu}\overline{\Theta}^{\mu\nu}
+2\sqrt{2\kappa}(V_{\mu\nu}{}^j A_j 
+\overline{\Theta}_{\mu\nu}\Theta) f^{\mu\nu}
+2\kappa (A_j A^j +\Theta^2)f_{\mu\nu}f^{\mu\nu}]
\]
\[
+
\frac{q^2}{64\pi^2f_a}\sqrt{-g}\epsilon^{\mu\nu\rho\sigma}
\!\Big(V_{\mu\nu}{}^j+\sqrt{2\kappa}(a_\nu D_\mu A^j-a_\mu D_\nu A^j
+A^j f_{\mu\nu})\Big)
\]
\[
\times
\Big(\Theta [V_{\rho\sigma j}+\sqrt{2\kappa}(a_\sigma D_\rho A_j
-a_\rho D_\sigma A_j
+A_j f_{\rho\sigma})]
+4(\overline{\Theta}_\rho +\sqrt{2\kappa}a_\rho\Theta) D_\sigma A_j\Big),
\]
and each fermion species of mass $M$ at the compactification scale contributes a term
\[
\frac{1}{\sqrt{-g}}{\cal L}_f=\overline{\psi} e^\mu{}_a\gamma^a
(i\partial_\mu +i\omega_{\mu}+qV_\mu)\psi
-M\exp\!\Big(-\sqrt{\frac{\kappa}{6}}\phi\Big)\overline{\psi}\psi.
\]
Again, as pointed out already in section \ref{kaka}, the graviphoton ensures
invariance of the effective four--dimensional gauge fields under diffeomorphisms
along the fifth dimension:
\[
V_\mu{}^j=A_\mu{}^j-\sqrt{2\kappa}a_\mu A^j
\]
\[
V_{\mu\nu}{}^j=
F_{\mu\nu}{}^j+\sqrt{2\kappa}(a_\mu D_\nu A^j-a_\nu D_\mu A^j-A^j f_{\mu\nu}).
\]

The classical equations of motion of the system above are invariant under constant shifts
of the dilaton
\begin{equation}\label{scale}
\phi\to\phi+c,
\end{equation}
\[
e^\mu{}_a\to\exp\!\Big(-\sqrt{\frac{\kappa}{6}}c\Big)e^\mu{}_a,
\]
\[
a_\mu\to\exp\!\Big(-\sqrt{\frac{2\kappa}{3}}c\Big)a_\mu,
\]
\[
V_\mu\to V_\mu, \qquad \overline{\Theta}_\mu\to\overline{\Theta}_\mu,
\]
\[
\psi\to\exp\!\Big(-\frac{1}{2}\sqrt{\frac{\kappa}{6}}c\Big)\psi,
\]
\[
A\to\exp\!\Big(\sqrt{\frac{2\kappa}{3}}c\Big)A,
\]
\[
\Theta\to\exp\!\Big(\sqrt{\frac{2\kappa}{3}}c\Big)\Theta,
\]
since this just rescales the action according to
\[
S\to\exp\!\Big(\sqrt{\frac{2\kappa}{3}}c\Big)S.
\]
This symmetry can equivalently be formulated as a scaling symmetry
on the co--ordinates, and the dilaton is often denoted as a Goldstone
boson for dilatations. However, the symmetry is unbroken as long as
the dilaton remains massless,
and I prefer the modern designation of the dilaton as a flat direction. 

The origin of the symmetry of the equations of motion
under (\ref{scale}) is easily
understood from the Kaluza--Klein origin of the action. The scale
transformations are equivalent to a rescaling of the internal dimensions
by a factor $\exp\!\Big(\sqrt{\frac{2\kappa}{3}}c\Big)$, and 
the equations
of motion resulting from a Kaluza--Klein Ansatz do not carry any remembrance
of the internal scale since we neglected any massive modes 
related to $\partial_5$.
Therefore, we should not expect a perturbatively generated dilaton
potential, since the perturbative
dynamics of the Kaluza--Klein zero  modes only
depends on the fluctuations of the internal dimensions through the
dilaton, but not on their actual size.
However, if there exist inherently four--dimensional
effects in the low energy dynamics, then we might expect
a non--perturbatively generated dilaton potential, since
unwinding of internal dimensions would certainly conflict
with inherently four--dimensional effects. A genuine four--dimensional
effect is the appearance of instantons in gauge theories, and
therefore we will concentrate on the issue whether instantons
create a dilaton potential.
Indeed, we will find that instantons create a dilaton
mass, because generically instantons imply that a
small dilaton would be energetically favored, while
the axions push the dilaton to large values.

This nicely complies with ideas about duality symmetries
between axions and dilatons: Instantons
create an effective axion potential and we have emphasized
before that there emerged much
evidence in recent years for a duality symmetry between
the axion and the dilaton, which is described in
eqs.\ (\ref{dualdec}--\ref{dual}) below.
A fundamental axion acquiring an effective potential
through instanton
mediated tunneling effects thus provides a very strong indication
for a light dilaton acquiring an effective potential
in a similiar vein.

  From a Minkowski space point of view an instanton contribution
to an effective dilaton potential may also be described as a gluon
condensate, and given the no--go conjecture for a perturbative
origin condensates provide a natural mechanism
to generate terms
in a dilaton potential.
Besides a gluon condensate we may expect from the coupling
to the kinetic energy of the axion a contribution from
a condensate $\langle\partial a\cdot\partial a\rangle$ \cite{rdhei,rdmpla3},
or from a gluino condensate \cite{DIN,DRSW,trt}, and we will take a 
very brief look
at a gluino condensate in the next section.
Recent discussions of contributions from gluino
condensates can be found in \cite{LNN}, where the coupling 
of the chiral
dilaton multiplet is re--examined, and in \cite{BGW}, where the dilaton
is treated in the linear multiplet. The proposal of a self--dual coupling
of the axidilaton to the gluons is reviewed in \cite{hpn}.

In the present section we will concentrate on the contribution from the axions and
its implications. Axions provide an attractive new mechanism
for dilaton stabilization since the exponents of the dilaton
multiplying the gluon and axion terms differ in sign, and since
non--trivial axion configurations provide suggestive explanations
for a condensate $\langle\partial a\cdot\partial a\rangle$.
This alternative proposal for generation of a dilaton potential
implies a drastic change of scales: While a gluino
condensate would be expected to generate a dilaton potential at
the SUSY breaking scale around or above 1 TeV, the axion would stabilize
the dilaton at the QCD scale around 1 GeV.

In the sequel the notation for the axion will be changed
 $\Theta\to a$, since graviphotons will be neglected
and $a$ is a more standard notation for the axion in four--dimensional
field theory. 
The main players in the game are then the dilaton $\phi$, the axion $a$
and gauge fields $A_\mu$ with field strengths $F_{\mu\nu}$, 
and their mutual interactions before taking into account 
non--perturbative
effects are governed by the Lagrangian
\begin{equation}\label{treeL}
\frac{1}{\sqrt{-g}}{\cal L}=\frac{1}{2\kappa}R-\frac{1}{2}g^{\mu\nu}
\partial_\mu\phi\cdot\partial_\nu\phi
-\frac{1}{2}\exp(-2\frac{\phi}{f_\phi})g^{\mu\nu}
\partial_\mu a\cdot\partial_\nu a
\end{equation}
\[
-\frac{1}{4}\exp(\frac{\phi}{f_\phi})
F_{\mu\nu}{}^j F^{\mu\nu}{}_j +
\frac{q^2}{64\pi^2f_a}\epsilon^{\mu\nu\rho\sigma} a F_{\mu\nu}{}^j
F_{\rho\sigma j}.
\]

The dilaton--axion and dilaton--gluon couplings in (\ref{treeL})
match in such a way that the system
exhibits an SL(2,$\mathbb R$)
duality symmetry, or S--duality for short, if the scales are related by
\begin{equation}\label{dualdec}
f_\phi= \frac{8\pi^2}{q^2}f_a.
\end{equation}
The invariance of the equations
of motion under the duality transformations
is most conveniently described in terms
of the axidilaton
\begin{equation}\label{axidil}
z=\frac{1}{f_a}[a+if_\phi\exp(\frac{\phi}{f_\phi})]
\end{equation}
and the symmetry is realized via
\begin{equation}\label{dual}
z'=\frac{a_{11}z+a_{12}}{a_{21}z+a_{22}},
\qquad\quad a_{11}a_{22}-a_{12}a_{21}=1,
\end{equation}
\[
F'_{\mu\nu}-i{\tilde{F}'}_{\mu\nu}
=(a_{21}z+a_{22})(F_{\mu\nu}-i\tilde{F}_{\mu\nu}),
\]
which means that the self--dual part of the Yang--Mills curvature
transforms like a half--differential on the axidilaton upper half--plane.

The invariance of equations of motion plus Bianchi identities
is easily recognized if the equations of motion are written as
\[
\frac{\partial^2 z}{(z-\bar{z})^2}-2\frac{\partial z\cdot\partial z}{(z-\bar{z})^3}
=\frac{f_a}{32if_\phi^3}(F+i\tilde{F})^2,
\]
\[
D_\mu\mbox{Im}\{z(F^{\mu\nu}{}_j-i\tilde{F}^{\mu\nu}{}_j)\}=0,
\]
where $\partial^2$ denotes the covariant Laplacian.

The scaling symmetry in (\ref{dual}) in the abelian case
is similar to but different from the
rescaling (\ref{scale}): $\delta a=2\varepsilon a$, $\delta \phi=2\varepsilon f_\phi$,
 $\delta A_\nu=-\varepsilon A_\nu$, and it implies a Noether current 
\begin{equation}\label{phicurr}
\frac{1}{\sqrt{-g}}j^\mu_\phi=f_\phi g^{\mu\nu}
\partial_\nu\phi +a\exp(-2\frac{\phi}{f_\phi})g^{\mu\nu}
\partial_\nu a-\frac{1}{2}\exp(\frac{\phi}{f_\phi})F^{\mu\nu}A_\nu
+\frac{q^2}{16\pi^2}\frac{a}{f_a}\tilde{F}^{\mu\nu}A_\nu.
\end{equation}

Both in abelian and non--abelian theories
the Peccei--Quinn symmetry $z\to z+a_{12}$ 
leaves the gauge potentials invariant and yields a conserved current
\begin{equation}\label{axicurr}
\frac{1}{\sqrt{-g}}j^\mu_a=f_a\exp(-2\frac{\phi}{f_\phi})g^{\mu\nu}
\partial_\nu a+\frac{q^2}{16\pi^2}\epsilon^{\mu\nu\rho\sigma}
(A_\nu{}^j\partial_\rho A_{\sigma j}+\frac{q}{3}f_{ijk}A_\nu{}^i A_\rho{}^j 
A_\sigma{}^k), 
\end{equation}
and
the scaling symmetry (\ref{scale}) of the equations of motion is also preserved
with $2\kappa\to 3f_\phi^{-2}$.

In view of the currents (\ref{phicurr},\ref{axicurr})  the designation of $f_\phi$ and $f_a$
as decay constants looks very natural and suggestive, 
since the scales parametrize non--vanishing matrix elements
between the (pseudo--)scalars and the vacuum, similar to the pion decay constant.
However, there is an important difference which should be kept in mind: 
The pion decay constant parametrizes
matrix elements $\langle 0|j_{\pi}^\mu|\pi\rangle$ which actually contribute to pion decays
into lepton pairs through intermediate vector bosons, and the matrix element arises
in the microscopic theory at low energies when the leptonic sector of the 4--Fermi--vertex
has already been evaluated. Nothing like that is expected to take place for a fundamental
axidilaton in string theory or Kaluza--Klein theory, and 
the scales $f_\phi$ and $f_a$ appear only with negative powers 
in physical matrix elements.

Yet we have not taken into account 
non--trivial field configurations
of the gauge fields and the axion:
We infer the non--perturbative effects of these field configurations 
from the Lagrangian of the
Euclidean action. In a flat background this takes the form:
\begin{equation}\label{treeLE}
{\cal L}_E=\frac{1}{2}g^{\mu\nu}
\partial_\mu\phi\cdot\partial_\nu\phi
+\frac{1}{2}\exp(-2\frac{\phi}{f_\phi})g^{\mu\nu}
\partial_\mu a\cdot\partial_\nu a
\end{equation}
\[
+\frac{1}{4}\exp(\frac{\phi}{f_\phi})
F_{\mu\nu}{}^j F^{\mu\nu}{}_j -i
\frac{q^2}{32\pi^2f_a}a \tilde{F}_{\mu\nu}{}^j
F^{\mu\nu}{}_j.
\]
Positivity of the real part and the estimate of the effective axion
potential by Vafa and Witten \cite{VW}
indicate that the dominating contributions to the path integral come
from instanton configurations $F=\pm\tilde{F}$ with constant dilaton
and the axion frozen to integer multiples of $2\pi f_a$.
This survival of instantons in the presence of the dilaton is crucial, since
integrality of the instanton number and invariance of the path integral
discretize Peccei--Quinn symmetry 
\[
\frac{a_{12}}{2\pi}\in\mathbb{Z},
\]
thereby also breaking
the scale invariance (\ref{scale}). 

The impact of instantons on the effective axion potential has been
examined by several authors, and the interpretation of instantons
as real time tunneling configurations between gauge theory vacua 
suggests 
\begin{equation}\label{axipot}
V(a) =m^2_a f^2_a\!\Big(1-\cos(\frac{a}{f_a})\Big)
\end{equation}
if the instanton gas
is dilute enough to neglect higher order cosine terms \cite{sc,GPY,jk}. 
While initially this result was inferred from semiclassical calculations
of tunneling amplitudes, the same potential can also be derived in a direct
instanton calculation if the wavelength of the axion is large compared
to the instanton size.

The picture emerging from this
shows us that
instantons create an effective axion potential with an enumerable
set of equidistant vacua, thus discretizing Peccei--Quinn
symmetry. Discreteness of the axion vacua and the cosine--like
shape of the axion in turn breaks the scale 
invariance (\ref{scale})
and indicates that the axidilaton--gluon system also lifts the degeneracy
of the dilaton. 
This is obvious in the gauge sector: Instantons
push the dilaton into the strong gauge coupling regime, since the action
of the instantons decreases with decreasing $\langle\phi\rangle$.
However, non--trivial configurations also arise in the axion sector:\\
-- If $a$ is periodic $a\sim a+2\pi f_a$,
then it contributes non--trivial configurations to the Euclidean
path integral over $\exp(-S_E)$ in the form of {\sl axion walls}
(instead of axion strings in three dimensions).
Periodicity of $a$ arises, if it is related to the argument
of a complex field with frozen modulus
in the low energy regime. This is e.g.\ the case if $a$ arises
as the phase of  a determinant of local fermion masses.\\
-- If $a$ is not an angular variable, then all the 
possible vacua $\langle a\rangle=2\pi f_an$ are distinct
and we expect three--dimensional domain walls separating four--dimensional
domains where $a$ approximates different vacua.

String theory is essential for the stability of these defects, since 
the four axion scattering
amplitude
at string tree level depends non--trivially on the momenta of the 
scattered axions \cite{jhsrep},
whence (\ref{treeL}) contains only
lowest order terms in a derivative expansion, as expected for an effective low
energy theory. However, we will not attempt a systematic derivation of the higher order
derivative terms from string scattering amplitudes, but rather adopt a phenomenological
approach in borrowing methods from the theory of cosmic strings to estimate
the axion condensate.

Both kinds of topological defects mark regions of non--vanishing
gradients $\partial a$ and favor large values of the dilaton
through the dilaton--axion coupling, thus compensating the effect of the
instantons.
Therefore, we expect an effective dilaton potential cutting off large values of the
dilaton through an average background field strength, while small values of the
dilaton are suppressed by the variance of the axion.
After adjustment of $q$ the potential results:
\begin{equation}\label{dilpom}
V(\phi)=
\frac{m_\phi^2 f_\phi^2}{6}\Big(2\exp(\frac{\phi}{f_\phi})
+\exp(-2\frac{\phi}{f_\phi})\Big).
\end{equation}
In the resulting model the dilaton mass $m_\phi$ and
the coupling constant $q$ parametrize the background field strength from
the instantons and the axion gradients from domain boundaries.

We may give estimates on $q$ and $m_\phi$ in terms of an average
instanton scale $\varrho$ and
a characteristic length $\Delta$ of the axion defects.
In the case of an angular axion $\Delta$ would measure the circumference
of the axion walls, while in the case of vacuum domains of the axion 
the four--dimensional domain boundaries are 
extended in three dimensions and have an
average
thickness $\Delta$ in the fourth direction.
The 
average separation $\varrho$ of instantons in the instanton liquid
is about three times larger than the average extension
of the instantons \cite{evs,ScSh}. From this we find an estimate
for the effective dilaton potential
\begin{equation}\label{dilpot1}
V(\phi)=\frac{16}{q^2\varrho^4}\exp(\frac{\phi}{f_\phi})
+2\pi^2\frac{f_a^2}{\Delta^2}
\exp(-2\frac{\phi}{f_\phi}).
\end{equation}
 This implies for the gauge coupling and the
dilaton mass
\begin{equation}\label{consist1}
\frac{1}{q}
=\frac{\pi f_a\varrho^2}{2\Delta},
\end{equation}
\begin{equation}\label{dilmass}
m_\phi f_\phi=\frac{4\sqrt{3}}{q\varrho^2}. 
\end{equation}

This investigation can be pursued further if $a$ is not an angular
variable: In this case we may estimate the parameter $\Delta$ by minimizing
the energy density of the axion domain boundaries
\begin{equation}\label{edensds}
u=2\pi^2\frac{f_a^2}{\Delta}+m_a^2 f_a^2\Delta
\end{equation}
yielding a thickness of the order
\begin{equation}\label{dsthick}
\Delta\simeq\frac{\pi\sqrt{2}}{m_a}
\end{equation}
which is of the same order as the thickness of ordinary axion domain walls
in Minkowski space \cite{KT}. From (\ref{consist1}) 
and (\ref{dsthick}) we find a relation between the axion
parameters and the average instanton radius
\begin{equation}\label{rela1}
m_a^2 f_a^2\simeq\frac{2}{\pi\alpha_q\varrho^4}.
\end{equation}

The average instanton radius at low temperatures is approximately \cite{evs}
\[
\frac{1}{\varrho}\simeq 200\, \mbox{MeV}
\]
and the big uncertainty in estimating axion parameters from (\ref{rela1})
concerns the coupling constant at the scale where instanton mediated tunneling
induces the axion mass. In QCD the investigation
of a truncated Dyson--Schwinger
equation for the gluon propagator
by Alkofer, Hauck and von Smekal
indicates  a value $\alpha_q\sim O(10)$  \cite{AHS},
and with the previous results this points at a value
\begin{equation}\label{estma}
m_a f_a\sim 10^{4}\, \mbox{MeV}^2.
\end{equation}

Although we did not use any information or assumptions
about axion--fermion couplings, it is remarkable that (\ref{estma})
complies with current algebra based estimates for an
axion coupling to light quarks \cite{BT,KSS,ms,GKR,BPY}:
\[
m_a f_a\sim m_{\pi}f_{\pi}.
\]

The potential (\ref{dilpot1}) and eq.\ (\ref{rela1}) imply a relation between the 
dilaton mass
and the axion mass
\begin{equation}\label{dilmass2}
m_\phi f_\phi\simeq\sqrt{6} m_a f_a.
\end{equation}
 
Since S--duality will certainly be broken at the scales
under consideration we 
expect a dilaton decay constant of the order of the Planck mass
while $f_a$ could be in the phenomenologically
preferred range $10^{10}-10^{12}$ GeV. Compared to the axion
this  implies
a much smaller mass of the dilaton and a later onset of coherent
dilaton oscillations.

\newpage
\section{The supersymmetric theory}\label{susyth}

The supersymmetric framework is of interest both from up--down and
bottom--up approaches to physics beyond the standard model of particle
physics. On the one hand on an intermediate energy scale
below the compactification scale string theory predicts
that the relevant physical degrees of freedom should be described
by a supergravity theory. On the other hand supersymmetric gauge theories
evolved into a primary tool for model building beyond the standard model,
and a footprint of supersymmetry is considered as one of the most
spectacular results that might be expected from accelerator physics
on foreseeable time scales. The subject got a further boost a few years
ago by the approximate convergence of coupling constants around $10^{16}$ GeV
if supersymmetry begins to apply at the TeV--scale\footnote{This analysis
is continuously improved, and predictions for masses of supersymmetric particles
are to a large degree model dependent \cite{BMP,LP}.} \cite{EKN,GKL,AdBF,LL}.

Given the necessity to introduce an axion to solve the strong CP problem,
supersymmetry provides an independent motivation to also introduce
a dilaton, since an axion in a chiral superfield always comes with
a dilaton \cite{Wilog,DIN2}.

With a few notational changes 
we will follow the conventions of Wess
and Bagger \cite{WB}. Supersymmetry is conveniently decribed in terms
of superfields. These are Grassmann valued fields over space--time, where
the Grassmann algebra is generated by a constant Dirac spinor $\theta$
of mass dimension $-\frac{1}{2}$:
\[
\{\theta_\alpha,\theta_\beta\}=0,\quad \{\theta_\alpha,\bar{\theta}_{\dot{\beta}}\}=0,
\quad \{\bar{\theta}_{\dot{\alpha}},\bar{\theta}_{\dot{\beta}}\}=0.
\] 
The particular superfields which we need are the chiral dilaton multiplet
\[
S=d+i\theta\cdot\sigma^\mu\cdot\bar{\theta}\,\partial_\mu d+\frac{1}{4}\theta^2
\bar{\theta}^2\partial^2 d+\sqrt{2}\theta\cdot\delta-\frac{i}{\sqrt{2}}\theta^2
\partial_\mu\delta\cdot\sigma^\mu\cdot\bar{\theta}+\theta^2 Z
\]
and the spinorial chiral superfield containing the gauge fields:
\[
W_\alpha=-i\lambda_\alpha+
\theta\cdot\sigma^\mu\cdot\bar{\theta }\,\partial_\mu\lambda_\alpha-\frac{i}{4} 
\theta^2 \bar{\theta}^2\partial^2  \lambda_\alpha+\theta_\alpha D
-\frac{i}{2}\theta^2\sigma^\mu{}_{\alpha\dot{\alpha}}\bar{\theta}^{\dot{\alpha}}
\partial_\mu D
\]
\[
+\frac{i}{2}(\sigma^\mu\cdot\bar{\sigma}^\nu)_\alpha{}^\beta\theta_\beta
F_{\mu\nu}+
\frac{1}{4}\theta^2
(\sigma^\mu\cdot\bar{\sigma}^\nu\cdot\sigma^\rho)_{\alpha\dot{\alpha}}
\bar{\theta}^{\dot{\alpha}}
\partial_\rho F_{\mu\nu}+\theta^2\sigma^\mu{}_{\alpha\dot{\alpha}}
(\partial_\mu\bar{\lambda}^{\dot{\alpha}}
-iq[A_\mu,\bar{\lambda}^{\dot{\alpha}}]).
\]
The lowest component $d$ of the dilaton multiplet is related to the 
axidilaton (\ref{axidil})
through
\[
d=i\frac{q^2}{16\pi^2}\bar{z},
\]
and we have normalized the multiplet such that it has mass dimension zero, i.e.\
the dilatino $\delta$ has mass dimension $\frac{1}{2}$. Note that our 
gauge field $A_\mu$
and the gauge fields $v_\mu$ in \cite{WB} differ in sign. The gluino $\lambda$
is the fermionic superpartner of the gluon.

The dilaton gluon coupling arises as the real part of the $\theta^2$--component of
 $SW^2$:
\begin{equation}
SW^\alpha_j W_\alpha^j|_{\theta^2}=-\lambda^\alpha_j\lambda_\alpha^j Z+
i\sqrt{2}\delta^\alpha\lambda_\alpha^j D_j-
\frac{1}{\sqrt{2}}(\sigma^\mu\cdot\bar{\sigma}^\nu)_\alpha{}^\beta 
F_{\mu\nu}{}^j\lambda^\alpha_j
\delta_\beta
\end{equation}
\[
+
d[D^j D_j+i(D_\mu\lambda)^j\cdot\sigma^\mu\cdot\bar{\lambda}_j
-i\lambda_j\cdot\sigma^\mu\cdot(D_\mu\bar{\lambda})^j-\frac{1}{2}F_{\mu\nu}{}^j
F^{\mu\nu}{}_j-\frac{i}{2}\tilde{F}_{\mu\nu}{}^j F^{\mu\nu}{}_j],
\]
and Witten observed that the correct supersymmetrization of kinetic terms 
of the axidilaton
is given by a logarithm of superfields \cite{Wilog}:
\begin{equation}
\ln(S+S^+)|_{\theta^2\bar{\theta^2}}=\frac{1}{(d+d^+)^2}(\partial d^+\cdot\partial d
+\frac{i}{2}\delta\cdot\sigma^\mu\cdot\partial_\mu\bar{\delta}
-\frac{i}{2}\partial_\mu\delta\cdot\sigma^\mu\cdot\bar{\delta}-Z^+Z).
\end{equation}
Neglecting the quarks and squarks, the supersymmetrization of (\ref{treeL})
is then given by
\begin{equation}\label{susyL}
{\cal L}=
-2f_\phi^2 \ln(S+S^+)|_{\theta^2\bar{\theta^2}}
+\frac{1}{2}SW^\alpha_j W_\alpha^j|_{\theta^2}
+\frac{1}{2}S^+\overline{W}{}^{\dot{\alpha}}_j 
\overline{W}{}_{\dot{\alpha}}^j|_{\bar{\theta}^2}
\end{equation}
if the decay constants satisfy the self--duality
condition (\ref{dualdec}).

In passing we also note that there exist
numerous possibilities to supersymmetrize the effective axion potential (\ref{axipot}),
e.g.\ through a superpotential
\[
{\cal V}=2m_a f_a^2 \cosh\!\Big(\frac{f_\phi}{f_a}(S-\frac{1}{2})\Big).
\]
However, since $m_a f_a\ll\Lambda_{SUSY}^2$ instantons will not dominate
the gauge theory vacuum
above the scale of supersymmetry breaking, and we will not pursue these superpotentials
further.

With eqs.\ (\ref{dilmass2}) and (\ref{susyL}) at hand
we may now also provide an estimate on the lower bound
of values of the dilaton decay constant for
which our approximation of dominance
of the axion condensate is applicable:

Elimination of the auxiliary field $Z$ of the dilaton multiplet yields the
dilaton gluon coupling
\[
{\cal L}_{\lambda\phi}=-\frac{1}{8f_\phi^2}\exp(2\frac{\phi}{f_\phi})\bar{\lambda}^2\lambda^2.
\]

If there is a gluino condensate with a scale $\Lambda_{SUSY}\sim 1$ TeV, 
this would contribute
a dilaton mass term\footnote{Note that restoration of the dilaton decay constant
shows that a dilaton mass from gluino condensates would be tiny, too. Therefore
the dilaton would also provide a suitable candidate for cold dark matter in the
framework of supersymmetry 
breaking.} $m_\phi^{\lambda}\simeq\Lambda_{SUSY}^3 f_\phi^{-2}$.
As a consequence the variance of the axion would dominate the dilaton mass at 
scales where
instanton induced tunneling becomes relevant for 
dilaton decay
constants above $10^{11}$ GeV, while for decay constant below this value the
gluino condensate would dominate to very low temperatures.

\newpage
\section{The dilaton as a dark matter candidate}\label{cosmo}

We have seen that both instanton tunneling and supersymmetry breaking
favor a very light weakly coupled dilaton accompanying a light weakly coupled
axion. However, light weakly coupled (pseudo--)scalars (or scalars for short)
are generically expected
to make an appreciable contribution to the energy density of the universe
in the form of cold dark matter, which in this specific instance means that
they developed coherent oscillations
with non--relativistic momenta. The onset of coherent oscillations is expected
when the universe has cooled down to temperatures where
mass terms begin to dominate over dissipative expansion terms in the equations
of motion of the scalars. Coherent oscillations are then
expected to dominate the energy density
of the scalars, since due to the weak coupling thermal creation and
annihilation of the scalars can be neglected. 

There exist several monographs where the general background for
cosmology and the impact of particle physics is very well
presented, see e.g.\ \cite{gb,KT,pjep}. However, I will begin with a review of a few
basic facts to set the stage for the discussion of the role the dilaton.

In discussing cosmological implications of the dilaton we will stick to the usual
approximation of spatial homogeneity and isotropy, i.e.\ we will discuss
dynamics in a Robertson--Walker space--time with line element
\begin{equation}\label{frwmet}
ds^2=-dt^2+{\cal R}^2(t)\Big(\frac{dr^2}{1-kr^2}+r^2d\vartheta^2+r^2\sin^2\vartheta
d\varphi^2\Big),
\end{equation}
where $r$ is 
dimensionless and the scale factor $\cal R$ has the dimension of a length. For $k=1$ 
the spatial part of the metric can be described as a 3--sphere 
of radius ${\cal R}(t)$
in flat $\mathbb{R}^4$, while $k=-1$ is the corresponding hyperbolic space. $k=0$
is flat 3--space. 

The form (\ref{frwmet}) of the metric implies that the matter energy momentum 
tensor has the form
 $T_{00}=\varrho(t)$, $T_{0j}=0$, $T_{ij}=p(t)g_{ij}$, and that energy 
conservation $\nabla_\mu T^{\mu 0}=0$ reads
\begin{equation}\label{econserv}
\frac{d}{dt}(\varrho{\cal R}^3)=-p\frac{d}{dt}({\cal R}^3).
\end{equation}
If particle interactions are fast enough to maintain thermal equilibrium
during expansion, the first and second law of thermodynamics
in a system with $N_i$ particles of chemical potential $\mu_i$
\[
dE=TdS-pdV+\sum_i\mu_i dN_i
\]
implies due to eq.\ (\ref{econserv})   
\begin{equation}\label{12haupt}
\frac{dS}{dt}=-\sum_i\frac{\mu_i}{T}\frac{dN_i}{dt},
\end{equation}
where $S$ and $N_i$ are the entropy and particle numbers in a unit
of comoving volume
 $v=V{\cal R}^{-3}$. 

I would like to add a remark on the thermodynamical expression for the
energy density appearing in the energy momentum tensor, since there
exists some confusion about this central object:

There are many ways to divide a thermodynamic potential by a volume, but
the energy density $T_{00}$ is
\[
\varrho=\frac{\partial E}{\partial V}\bigg|_{t}.
\]
However, in a FRW universe temperatures and chemical potentials will
only depend on $t$, while particle numbers also go with the volume.
 From this we may immediately translate the expression for $\varrho$
into thermodynamics:
\[
\varrho=\frac{\partial E}{\partial V}\bigg|_{T,\mu}=
T\frac{\partial S}{\partial V}\bigg|_{T,\mu}-p
+\mu\cdot\frac{\partial N}{\partial V}\bigg|_{T,\mu},
\]
with an obvious abbreviation for the sum over particle species. From the grand potential
 $\Omega_{GC}(T,V,\mu)=-pV$, $d\Omega_{GC}=-SdT-pdV-N\cdot d\mu$
we learn that
\[
\frac{\partial S}{\partial V}\bigg|_{T,\mu}=\frac{\partial p}{\partial T}\bigg|_{V,\mu},
\]
and that the particle densities are
\[
\nu_i\equiv\frac{\partial N_i}{\partial V}\bigg|_{T,\mu}=
\frac{\partial p}{\partial \mu_i}\bigg|_{T,V,\hat{\mu}_i},
\]
where $\hat{\mu}_i=\mu_1,\ldots,\mu_{i-1},\mu_{i+1},\ldots$.
With $\beta=\frac{1}{kT}$ a useful expression for $\varrho$ is then
\begin{equation}\label{therdic}
\varrho=T\frac{\partial p}{\partial T}\bigg|_{V,\mu}-p+\mu\cdot\nu
=\mu\cdot\nu-\frac{\partial}{\partial\beta}(\beta p)\bigg|_{V,\mu},
\end{equation}
because this directly relates $\varrho$ to $\Omega_{GC}$.

In the case of one particle species the 
dynamical evolution would then be determined as follows:
The energy levels 
of the particles determine the grand potential and the pressure $p(T,\mu)$.
 Eq.\ (\ref{therdic}) is then used to calculate $\varrho(T,\mu)$, while in the thermodynamic
limit the entropy and particle
number in a comoving volume are $S=\frac{\partial p}{\partial T}{\cal R}^3$,
 $N=\frac{\partial p}{\partial\mu}{\cal R}^3$.
Equations (\ref{econserv},\ref{12haupt})
and the Friedmann equation (\ref{friedeq}) below
then constitute a set of three first order differential equations
for the dynamical variables ${\cal R}(t)$, $T(t)$ and $\mu(t)$

The algebra of dynamical degrees of freedom in
a quasi--statically expanding FRW model is therefore
generated by
the scale parameter which describes expansion or contraction 
with a relative velocity $H=\frac{d}{dt}\ln({\cal R})$, the temperature $T$
and the chemical potentials of the various
particle species. 
Of course, in practice equilibrium is not maintained for weakly interacting
particle species 
which decouple due to thinning out in the expanding universe. This is taken care of by
assigning extra effective temperatures governing the energy distribution
within these particles species.
 Furthermore, it proved a very useful approximation to calculate
the history of the homogeneous and isotropic background piecing together
different epochs where the energy content of the universe was stored primarily
either in radiation, in pressureless matter, or in scalar fields. 
In the first two cases the relevant
degrees of freedom for the evolution
of the background metric are $\varrho$, $p$ and $\cal R$, and eq.\ (\ref{12haupt})
is replaced by a dispersion relation $p=p(\varrho)$,
while in the case of scalar fields the relevant degrees
of freedom are the scalar fields and  $\cal R$, and the evolution is governed by the
equations of motion of the scalar fields and the Friedmann equation.
The explicit matter content
and interactions determine the sequence and transitions of epochs in this
approximation, and the present epoch of dust dominance $p\simeq 0$ outnumbers
all previous epochs since expansion
of the very hot and dense primeval plasma
in its duration $\sim 10^{17}$ seconds. When solutions of the Friedmann
equation for $p>-\varrho$ are evolved backwards in time we inavoidably hit
an initial singularity $\varrho\to\infty$, ${\cal R}=0$ for finite parameter $t$,
and we will stick to the
usual terminology of initial singularity or big bang, although we can only be sure
that
there existed a very hot and dense phase at some very early stage of our
contemporary epoch\footnote{This qualification may seem strange, since our present
investigation is mainly motivated from string theory. However, even within string theory
we are currently in a phase of exploring more and more hitherto unknown
(and unexpected)
possibilities for the high energy sector of the theory, and yet there has not emerged
a coherent proposal for the evolution of space--time near the Planck scale.

Besides this, quantum groups provide another set of challenging ideas about the
shortest distance structure of space--time.}.
The Hubble parameter $H_0$
corresponding to the present value of $H(t)$ is still subject to a seminal debate
among astronomers, and this is encoded in an uncertainty parameter $h$
which varies between $0.5< h< 0.85$ \cite{RPP}:
\[
H_0=100h\frac{\mbox{km}}{\mbox{Mpc}\cdot \mbox{s}}=1.02h\times 10^{-10}\frac{1}{\mbox{yr}}.
\]

It may be worthwhile to recall for the justification of (\ref{frwmet}) 
that the DMR experiment on COBE
measured temperature anisotropies in the cosmic microwave background of order
 $\frac{\delta T}{T}\simeq 10^{-5}$ in multipole expansions up to $l=30$, 
see \cite{COBE1,COBE2,kmg}
and references there, as well as \cite{SSW} and \cite{LBBB} for compilations
of measurements since COBE. 
The exciting result for the experts was the 
transition from an era of upper bounds on the
anisotropy to actual measurements, 
but the results also show how good an approximation an isotropic universe
represents up to
the decoupling of the cosmic background radiation around  $10^{12}$
seconds after the big bang\footnote{Homogeneity is much more subtle from the experimental
point of view, see the discussion in \cite{gb}.}.

Taking into account energy conservation (\ref{econserv}),
the Einstein equations reduce to
\begin{equation}\label{friedeq}
\dot{\cal R}^2+k=\frac{\kappa}{3}{\cal R}^2\varrho
\end{equation}
and a flat universe would correspond to an averaged contemporary mass density
\[
\varrho_c=\frac{3}{\kappa}H_0^2=
1.9h^2\times 10^{-26}\frac{\mbox{kg}}{\,\mbox{m}^3}=81h^2(\mbox{meV})^4
=2.4h^2\times 10^{-120}m_{Pl}^4.
\]

 Following the usual habit among astronomers energy densities will be measured
in units of $\varrho_c$ in this section: $\Omega=\frac{\varrho}{\varrho_c}$.

Particle physics influences (and is increasingly influenced by) cosmology in many ways.
Two instances where scalar particles play major roles are inflation and dark matter:

Inflation denotes a phase of accelerated expansion of the universe when distances grew
faster than light cones. This happens for pressure to density ratios 
between $-1\leq\frac{p}{\varrho}<-\frac{1}{3}$, and temporary superluminal expansion
has the potential to solve several
major problems in cosmology, in particular the horizon problem, the neglegibility
of contributions to $\Omega$ from
topological defects, and the problem
why the measured energy density is not far away from $\varrho_c$. 
These and other motivations for inflation are very thoroughly 
reviewed in \cite{gb,KT,pjep}. 

Although inflation seems to become an integral part of ongoing
extensions of 
the standard cosmological model, I will
not address it any further, since I do not expect that the dilaton
which we examine here provides a suitable candidate
for the sought for inflaton:
We will see that thermally produced dilatons contribute at most 2 percent
to the energy density of the universe at the scales where instantons
and axions are expected to induce a mass term. On the other hand, 
this mass term also implies that the dilaton field can store energy 
in coherent oscillations which can provide a considerable amount
of the contemporary energy density. Contrary to radiatively stored energy
this energy would only very slowly dissipate into thermal energy of non--relativistic
matter.
In discussing this we will rely on the conservative assumption that at the time
of onset of oscillations most energy is still stored in relativistic matter, i.e.\
we will calculate in a given Friedmann--Robertson--Walker background
expanding according to ${\cal R}\sim\sqrt{t}$.
Then the oscillations behave like non--relativistic matter with their energy density
decreasing according to $\varrho\sim{\cal R}^{-3}$. In an alternative scenario
one might speculate that
the axion or dilaton could trigger temporary
superluminal expansion of the universe with
approximately constant energy density, if the axion and/or dilaton field
would dominate the energy density immediately after the onset of oscillations.
However, this would require a yet unknown mechanism to convert most of the
thermal energy in relativistic matter into coherent axion or dilaton fields,
and it would conflict with the successful interpretation of the cosmic microwave
background as a remnant of the hot radiatively dominated phase before
recombination if the energy is not re--converted into radiation.

Contrary to the verdict about dilaton induced inflation in FRW backgrounds, 
it turns out that a light dilaton can very well contribute to dark matter
in the universe:
There is wide agreement in the astronomy/astrophysics community that a considerable
fraction of the contemporary
energy density of the universe must be due to non--luminous matter,
and the problem is to determine the nature of this matter. 
It seems clear now that part of this matter is of non--baryonic origin, since even for the
lowest possible values of Deuterium abundance\footnote{Low D abundance
means large conversion into ${}^4\mbox{He}$ due to large baryon density.} 
and a Hubble constant as low 
as $50\frac{\mbox{\footnotesize km}}{\mbox{\footnotesize Mpc}\cdot \mbox{\footnotesize s}}$ 
primordial nucleosynthesis
allows for a maximal
baryonic contribution to the energy density of order $\Omega_B\leq 0.08$ \cite{mst2},
while both galactic motion and velocities on larger scales indicate values of $\Omega$
beyond $0.1$.
The recent survey of large scale peculiar motion in the nearby universe
of Strauss and Willick \cite{StWi}
gives a range $0.3\leq\Omega\leq 1$. Large scale motion, structure formation
through gravitational attraction, and gravitational lensing
also indicate that a considerable
amount of dark matter is concentrated in halos around galaxies or groups of galaxies,
whence a large fraction of the dark matter has to be cold. For very massive
particles this means that they had to be non--relativistic when they decoupled
due to thinning out in the expanding universe. Very weakly interacting
light particles contribute to cold dark matter (CDM) through non--relativistic
coherent oscillations, as has been pointed out before and will be explained below. From
the particle physics point of view
the leading contenders for the CDM component of dark matter are the axion \cite{KT,jk},
the lightest supersymmetric particle commonly
denoted as a neutralino, and more recently the dilaton \cite{GV,DV,rdhei,rdmpla3,mg}.

In this section
we will focus on a dilaton whose mass at very low temperatures
is dominated through instanton effects and discuss 
evolution of the axidilaton in an expanding universe. 
 For this purpose we will concentrate on a temperature range
between 1 TeV and 100 MeV, since we expect that a light axidilaton
acquires its masses in that 
range  (for $10^{11}\mbox{GeV}\leq f_\phi\leq 10^{18}$GeV) and it makes 
sense to assume
that besides our hypothesized axidilaton no further
degrees of freedom beyond the standard model will be relevant
at these scales. 

The universe has cooled down to temperatures 1 TeV and 100 MeV
around $10^{-13}-10^{-12}$ and $10^{-5}-10^{-4}$
seconds after the initial singularity\footnote{The
robustness of these scales against our ignorance of
particle physics at very high energies is 
amazing: Eqs.\ (\ref{tTrel},\ref{tTrel2}) show that
supersymmetry would divide these time scales only by a factor $\sqrt{2}$,
and that one would need $10^4$ additional relativistic degrees of freedom
to invalidate the order of magnitude estimates!}. In this energy range 
the universe is radiation dominated with relativistic background matter
satisfying a dispersion relation $\varrho=3p$. The density 
and the scale factor then evolve according to 
\begin{equation}\label{tTrel}
\varrho(t)=\frac{3}{4\kappa t^2}
\frac{({\cal R}_0^2+kt_0^2)^2}{({\cal R}_0^2+kt_0(t_0-t))^2},
\end{equation}
\begin{equation}\label{scalevolv}
{\cal R}(t)=\sqrt{\frac{t}{t_0}}\sqrt{{\cal R}_0^2+kt_0(t_0-t)},
\end{equation}
where ${\cal R}_0$ is the scale parameter at a fixed time $t_0$ during radiation
dominance. Evolving back the current energy density, which is within one order
of magnitude of the critical density, shows that ${\cal R}_0\gg kt_0$ during radiation
dominance and curvature effects can be neglected.
The energy density can then be estimated to relate time and temperature 
scales\footnote{Without approximate flatness we would find a curvature
term $-k{\cal R}^{-2}(P_i X^i)^2$ in the dispersion relation, and even the 
treatment of ideal gases would become very complicated.}:
\begin{equation}\label{tTrel2}
\varrho=\frac{\,\pi^2}{30}g(T)T^4,
\end{equation}
where $g(T)$ is the effective number of relativistic degrees of freedom in thermal
equilibrium and varies by a factor of two for temperatures between
 100 MeV and 1 TeV: If we assume standard model particle content plus an axidilaton
we find in the high temperature regime $g(1\mbox{TeV})=105.75$ including the 
Higgs particle\footnote{Right--handed neutrinos
were excluded in the calculation of $g(T)$.} 
and
the top quark, or  $g(1\mbox{TeV})=94.25$ without them. In the low temperature sector we find
 $g(1\mbox{GeV})=g(\mbox{100MeV})=49.75$, 
where the light particles included are the electron, 
up and down quarks, three left--handed neutrinos, the photon, eight gluons, the axion
and the dilaton. 

Thermally
produced dilatons
are relativistic, and their equilibrium density at temperature $T$ is:
\[
\nu(T)=\frac{\zeta(3)}{\pi^2}T^3.
\]
 From this and eqs.\ (\ref{tTrel}--\ref{tTrel2}) we find a mild increase
of the number of dilatons
per comoving volume $v=V{\cal R}^{-3}$ with temperature:
\[
N_\phi =\frac{\zeta(3)}{\pi^2}\Big(\frac{45}{2g(T)}\Big)^{3/4}
\Big(\frac{m_{Pl}}{\pi t_0}\Big)^{3/2}{\cal R}_0^3.
\]
This can easily be understood: As temperature approaches mass thresholds 
the effective number of 
relativistic degrees of freedom decreases and particles annihilate into
the remaining light
degrees of freedom.

We learn from $g(T)$ that thermally created
dilatons contributed about 2\% to the energy density of 
the universe for 1 GeV$>T>$100 MeV, if the dilaton was still in thermal
equilibrium. On the other hand, 
we have seen that the dilaton is extremely weakly coupled, and it may
well happen that it decouples from the heat bath at a temperature $T_{dec}$
above 1 GeV.
Then the energy density of dilatons which were thermally produced 
 at the temperature $T_{dec}$ is still governed by the
temperature $T$ of the heat bath as long as the heat bath remains relativistic.
This holds for any massless decoupled particle species and is a simple
consequence of the fact that the energy density of thermally produced decoupled species
evolves according to $\varrho_{dec}\sim {\cal R}^{-4}\sim T^4$, i.e.\ the decoupled species
cools out exactly like the relativistic heat bath.
There is a difference, of course: The number of effective relativistic degrees of freedom
seen by the decoupled particles was $g(T_{dec})>g(T)$, and if thermally produced
dilatons decoupled at $T_{dec}$ their contribution to the energy density
of the universe was $g(T_{dec})^{-1}<2$\%, while their number density was reduced 
by a factor $(g(T)/g(T_{dec}))^{3/4}$. After photon recombination the contribution
of thermal dilatons to $\Omega$ becomes negligible since $\varrho_{\phi,th}$
still decays with ${\cal R}^{-4}$, while the energy density in the 
dust decays only with ${\cal R}^{-3}$.
Therefore, with or without decoupling thermally produced dilatons make no
relevant contribution to the present energy density of the universe.

How then do the worries arise that
the coupling scale of the dilaton is constrained to values below $10^{12}$ GeV
 from the requirement $\Omega\leq 1$,
similar to the decay constant of the axion? This follows from the
corresponding analysis for the axion in \cite{PWW,AS,DF,mst}, 
if $f_\phi$ is supposed to determine the expectation value $\langle\phi^2\rangle$
of the dilaton at the onset of coherent oscillations.
However, 
there is a caveat in this reasoning: It follows from the form of the instanton
induced potential for the axion that its amplitude at the onset of oscillations
is of the order $f_a$, since both a periodic and a non--periodic axion is at most
 $|\Delta a|\leq\pi f_a$ away from a local minimum of $V(a)$.
On the other hand, such an estimate makes no sense for the dilaton
since the low energy potential is not periodic and we have encoded any 
non--vanishing expectation value
of the dilaton in our horizon in the gauge coupling.
We also should not rely on dimensional arguments, since at the onset of oscillations
there are two widely different mass scales which govern the dynamics of the
dilaton: A very large decay constant and a very small mass.

As a consequence, we employ the relation (\ref{dilmass2})
\[
m_\phi f_\phi\simeq\sqrt{6} m_a f_a
\]
 to estimate the contribution
of dilaton oscillations to $\Omega$.
Coherent oscillations of the axion and the dilaton arise when
the mass terms begin to dominate over the expansion terms in the 
equations
of motion of scalars:
\begin{equation}\label{eqmotscal}
\ddot{\phi}+\frac{3}{2t}\dot{\phi}+m^2\phi=0.
\end{equation}
Classical trajectories of the axion and the dilaton satisfy this equation
approximately, since due to the large decay constants
the axion--dilaton coupling and the couplings
to gauge fields provide negligible corrections to the linearized theory.

As long as mass terms can be neglected
scalar fields approach stationary values with deviations fading with $t^{-1/2}$.
On the other hand the field oscillates with frequency $m$ if the mass term
dominates, and this misalignment mechanism promotes light scalars to cold dark
matter even though the temperature exceeds the mass.
As a very heuristic argument to identify the transition region 
between the two regimes one may require
smooth transition of $\dot{\phi}$. This gives for the transition time $2\tilde{t}=3m^{-1}$.
 For times $t\gg m^{-1}$
the energy density $\varrho_\phi=\frac{1}{2}\dot{\phi}^2+\frac{1}{2}m^2\phi^2$
stored in the oscillations behaves exactly like
pressureless matter under expansion:
Dine and Fischler \cite{DF} 
pointed out that the constant mass solution to (\ref{eqmotscal})
is 
\[
\phi(t)=t^{-\frac{1}{4}}(A_+ J_{\frac{1}{4}}(mt) +A_- J_{-\frac{1}{4}}(mt)),
\]
and the asymptotic expansion for $mt\gg 1$ implies $\varrho_\phi\sim{\cal R}^{-3}$.
This scaling behavior of $\varrho_\phi$ persists in a dust dominated universe, where
 $\phi$ evolves with $t^{-1/2}J_{\pm\frac{1}{2}}(mt)$
and $\cal R$ evolves with $t^{2/3}$, and the oscillations do not 
contribute to the pressure in the universe.
Several other groups have analyzed the influence of the instanton induced axion 
mass
and found that it begins to dominate over the expansion term
at $T_a\simeq 1\mbox{GeV}$ \cite{PWW,AS,mst}.

A calculation of the 
temperature dependence of the axion mass by Gross, Pisarski and Yaffe
was employed by Turner to determine the mass dependence of the temperature $T_a$
beyond which the evolution of axions
is dominated by oscillatory behavior \cite{mst}. He found that the transition
temperature scales with $m_a$ according to $T_a\sim m_a^{0.18}$, 
where $m_a$ is the low temperature limit of the axion mass.
Due to the constancy of
the decay constants Turner's result also applies to the dilaton
and we find for the corresponding
scale 
\[
\frac{T_\phi}{T_a}\simeq\Big(\frac{f_a}{f_\phi}\Big)^{0.18},
\]
whence the momenta of the oscillations entering the horizon are related by
\[
\frac{p_\phi}{p_a}\simeq\frac{t_a}{t_\phi}\simeq\Big(\frac{f_a}{f_\phi}\Big)^{0.36}.
\]

 From these ratios follows an estimate on the velocity ratio, which can be
used as a further indicator that coherent dilaton oscillations
qualify as cold dark matter:
\[
\frac{v_\phi}{v_a}\simeq\Big(\frac{f_\phi}{f_a}\Big)^{0.64}.
\]

To discuss implications of the previous results 
on the role of the dilaton in
cosmology we should distinguish two cases:\\
-- S--duality applies at the GUT scale, implying that $f_\phi$ is bounded to 
be at most two orders of magnitude above the
expected value for a misaligned axion $f_a\simeq 10^{12}$ GeV.\\
-- S--duality is broken, with $f_a\simeq 10^{12}$ GeV 
but $f_\phi\simeq 10^{18}$ GeV.

The consequences in the first case
are schematically similar to the consequences in the second case,
but it leaves us with the puzzle
to identify a mechanism which could lower $f_\phi$ by four orders of magnitude
from its theoretically expected value. A further case very similar to the case
of S--duality at the GUT scale would suppose S--duality at the QCD scale.
A priori there seems no particular justification for this assumption, apart from the
fact that it nicely complies with the mass estimate $m_\phi\geq 10^{-4}$ eV which would arise
for a dilaton coupling to nucleon masses \cite{EKOW}. Then the dilaton decay constant
would be close to the axion decay constant and
 the axion and the dilaton would 
develop oscillations at the same scale and make comparable contributions to $\Omega$.

In the second case 
S--duality is maintained in the axidilaton sector, but not in the couplings to gauge fields.
Eq.\ (\ref{dilmass2}) then
hints at a non--perturbatively generated dilaton mass which is 
much smaller
than the axion mass:
\[
m_\phi\sim 10^{-6}m_a
\]
and
the dilaton will start to oscillate after the axion, 
when the temperature
has dropped by another factor of 10 and the time scale has expanded by two orders
of magnitude.
 
The velocity of large scale
fluctuations entering the horizon at the
QCD scale is $v_\phi\sim 10^4 v_a$, and from $v_a\sim 10^{-6}$ \cite{gb} 
we learn that even in this sense dilaton oscillations
remain non--relativistic for all choices of $f_\phi$.
Borrowing on the results of \cite{PWW,AS,DF,mst}
for the axion we find for the dilaton 
contribution to the
contemporary energy density of the universe
\begin{equation}\label{omega}
\frac{\Omega_\phi}{\Omega_a}
\simeq 10^{-5}\frac{\langle\phi^2\rangle}{f_a^2}\simeq
10^7\frac{\langle\phi^2\rangle}{m_{Pl}^2},
\end{equation}
and the dilaton would make an appreciable contribution to
the energy density for misalignment in the range
\[
\frac{\sqrt{\langle\phi^2\rangle}}{\,m_{Pl}}\sim 10^{-3}-10^{-4}.
\]
Taking into account the time evolution of the dilaton
before instanton tunneling this
 corresponds to 
\[
\frac{\sqrt{\langle\phi^2\rangle}}{\,m_{Pl}}\simeq 1
\]
for temperatures above 1 TeV, and the outlook for an appreciable
fraction of dilatons in the dark matter 
seems promising. However, any further investigation of this subject
requires better knowledge, or speculation, about new physics
and evolution of a massless dilaton
for temperatures above 1 TeV, 
and this is beyond the scope of the present work.

\newpage
\section{Conclusions and outlook}\label{conc}

The appearance of fundamental scalars with a direct
coupling to gauge curvature terms remains a
challenge in string theory which offers unexpected
rewards in low energy physics.

In order to resolve the ambiguity in the definition
of gauge couplings in the presence of a massless dilaton,
the dilaton has to acquire a mass at an early stage in the
evolution of the universe. Motivated from the observation
that both in string theory and in Kaluza--Klein theory
the dilaton couples with different signs to axions and
to gluons the proposal was made that rather than a gluino
condensate it is a variance of the axion in the Euclidean domain
which stabilizes the dilaton. For consistency this proposal
has to rely on the assumption that the four--dimensional
field theory containing axions and dilatons is an effective
theory, with topological defects stabilized through higher
derivative terms, as is the case e.g.\ in string theory.
 An S--dual coupling between the axion and the dilaton
then yields an estimate on the dilaton mass
 $m_\phi\simeq m_a f_a f_\phi{}^{-1}$. Comparison with
the simplest supersymmetric extension of an axidilaton--gluon
theory revealed that the axion coupling should dominate
the dilaton mass for decay constants $f_\phi>10^{11}$ GeV
and gluino condensates below 1 $\mbox{TeV}^3$.

We have pointed out that the dynamics of a light
scalar in an expanding universe before mass dominance
easily accomodates for large coupling scales without
overclosing the universe as long as no multivalued vacua
emerge. For a dilaton with coupling scale $f_\phi\simeq m_{Pl}$
this means that a variance $\sqrt{\langle\phi^2\rangle}\simeq m_{Pl}$
is still permissible at a temperature $\simeq 10^3 T_\phi$, where
 $T_\phi$ is the temperature where coherent dilaton oscillations evolve.
In this case we have seen that the dilaton provides an interesting
candidate for cold dark matter accompanying an axionic component,
and for a coupling to QCD we found an estimate $T_\phi\simeq 10^{-1}T_a$.
As a consequence
the onset of dilaton oscillations seems to
be close to or even coincide with the QCD phase
transition.

We have also seen that a dilaton coupling to gauge curvature terms
provides a simple mechanism to accomodate both a regularized
Coulomb potential and a confining potential in gauge theory.
Given this observation and the fact that string theory
unavoidably predicts a dilaton coupling
to gauge fields, continuing
investigation of light dilatonic degrees of freedom
seems more than justified. It is of particular interest to see
how the transition from the confining solution
to a regularized Coulomb potential proceeds, and which parameters
control the phases of a gauge theory coupling to a dilaton.

In conclusion, gauge theories with a dilaton present rather an interesting
than a worrisome prediction of string theory.

\newpage
\section*{Appendix A: Conventions and notation}

We use Greek letters
for three-- and four--dimensional holonomic indices, while
anholonomic indices are denoted by Latin letters from the
beginning of the alphabet. 
Higher dimensional holonomic tangent frame indices are denoted by capital letters
from the middle of the alphabet and anholonomic indices by capital letters from the
beginning of the alphabet. Hence, components of the $4$--bein 
and the $D$--bein in $D>4$
dimensions read $e_\mu{}^a$, $E_M{}^A$.
Latin letters from the middle of the alphabet are used both for Lie algebra indices
and for spatial Minkowski space indices in $3+1$ dimensions, and matrix elements
of Lie algebra generators are written as $(X_i)_{ab}$.
We use a boldface notation for 3--vectors.
Gauge couplings
are usually denoted by $q$, 
while $g$ is reserved for the metric in three or four dimensions. 

Our conventions for Planck units are rescaled by a factor $\sqrt{8\pi}$:
\[
m_{Pl}=\kappa^{-1/2}=(8\pi G)^{-1/2}=2.4\times10^{18}\,\mbox{GeV}, 
\]
\[
t_{Pl}=2.7\times 10^{-43}\,\mbox{s},
\]
\[
l_{Pl}=8.1\times 10^{-35}\,\mbox{m}.
\]
In the literature these units are sometimes referred to as reduced Planck units.

The generic setting for quantum field theory are total spaces fibered
by a usually highly reducible representation space of a 
group $SO(1,D-1)\times G$.
$G$ is referred to as a gauge group, and
is assumed to be a compact Lie group consisting of simple factors. It is generated
by a Lie algebra
with relations
\[
[X_i,X_j]=if_{ij}{}^k X_k.
\] 

The fiber projects down to a $D$--dimensional
base space $\cal M$
of Minkowski signature $(-,+,\ldots,+)$, and $SO(1,D-1)$
is the structure group of the tangent bundle.
The generators of $SO(1,D-1)$ as well as their representations
are denote by $L_{ab}=-L_{ba}$.

Covariant derivatives and curvatures are defined via
\[
D_\mu=\partial_\mu+\omega_\mu-iqA_\mu=
\partial_\mu-\frac{1}{2}\omega^a{}_{b\mu}L^b{}_a-iqA_\mu{}^j X_j,
\]
\[
R_{\mu\nu}-iqF_{\mu\nu}=-\frac{1}{2}R^a{}_{b\mu\nu}L^b{}_a-iqF_{\mu\nu}{}^j X_j
=[D_\mu,D_\nu],
\]  
and the dual curvature tensor $\tilde{F}$ in four dimensions is 
\[
\tilde{F}_{\mu\nu}= \frac{1}{2}\epsilon_{\mu\nu\rho\sigma}F^{\rho\sigma}.
\]
 $F$ is denoted as self--dual in Minkowski space if
\[
\tilde{F}_{\mu\nu}=iF_{\mu\nu}.
\]

A
Wick rotation maps the Schr\"odinger equation
into the diffusion equation through\footnote{van Nieuwenhuizen and Waldron
recently pointed out that the Wick rotation has a continuous extension
in terms of a five--dimensional Lorentz--boost. They employed
this observation to identify
the action of the Wick rotation on spinors \cite{vNW}.}
\[
t\to -i\tau
\]
and the transition from Minkowski
space field theory to Euclidean field theory proceeds via
\[
\varphi(t,\vek{x})\to i^{\#(0)}\varphi_E(\tau,\vek{x}),
\]
\[
{\cal L}(\varphi)\to -{\cal L}_E(i^{\#(0)}\varphi_E),
\]
where $\#(0)$ denotes the net number of covariant timelike indices
in the field $\varphi$. The $\epsilon$--tensor is not covariantly transformed
under Wick rotation and satisfies
 $\epsilon^{0123}=\frac{1}{\sqrt{|g|}}$, thus accounting for the oscillatory instanton
contribution through the axion--gluon coupling.

On the level of partition functions
\[
Z[J]=\exp(iW[J])=\int D\varphi\exp(iS[\varphi]+i\int d^4x J\cdot\varphi)
\]
is mapped to
\[
Z_E[J_E]=\exp(-W_E[J_E])=\int D\varphi_E\exp(-S_E[\varphi_E]+\int d^4xJ_E\cdot\varphi_E).
\]
The mean fields are
\[
\phi(x)=\frac{\delta W[J]}{\delta J(x)},
\]
\[
\phi_E(x)=-\frac{\delta W_E[J_E]}{\delta J_E(x)},
\]
and the effective actions are accordingly
\[
\Gamma[\phi]=W[J]-\int d^4x J(x)\cdot\phi(x),
\]
\[
\Gamma_E[\phi_E]=W_E[J_E]+\int d^4x J_E(x)\cdot\phi_E(x).
\]
The effective actions and the mean fields thus encode the quantum dynamics
of the system under consideration in terms of classical evolution equations:
\[
\frac{\delta\Gamma[\phi]}{\delta\phi(x)}=-J(x),
\]
\[
\frac{\delta\Gamma_E[\phi_E]}{\delta\phi_E(x)}=J_E(x).
\]
Note that under Wick rotation $\delta(x)\to i\delta(x)$, and 
therefore $J(t,\vek{x})\to (-i)^{\#(0)}J_E(\tau,\vek{x})$, but
\[
\frac{\delta}{\delta J(x)}\to i^{\#(0)+1}\frac{\delta}{\delta J_E(x)}.
\]
The mean fields thus transform like the 
quantum fields: $\phi(x)\to i^{\#(0)}\phi_E(x)$.

\newpage
\section*{Appendix B: A remark on perturbative aspects
        of the axidilaton}

We have argued from the Kaluza--Klein type coupling of a dilaton to gauge
fields that instantons should provide a mechanism to stabilize a dilaton
in four dimensions, irrespective from the presence or absence
of higher massive modes. It is tempting to push this argument a little further
and conclude that no effective potential should be generated perturbatively:
In a low--dimensional Kaluza--Klein theory the volume of internal dimensions
effectively only rescales the higher--dimensional Planck mass and axion constant 
to their
low--dimensional values, and we have seen that the axion and
dilaton couplings
carry no further remembrance of the compactification scale. Thus
in the low energy approximation
the compactification scale reduces to a mass threshold, but there is no
traceable imprint of this scale in the low energy sector\footnote{Of course, in 
a real Kaluza--Klein system there would be scalars in the adjoint representation
of the gauge group, and with chiral fermions one could infer the likely
existence of a compactification scale, but 
one would not have any
hint for its order of magnitude.}. Hence perturbation theory in 
the low energy
sector can not reveal the presence of a compactification scale or its actual value,
and therefore it can not remove the degeneracy of the dilaton.
The same conclusion then
should apply to any theory with a dilaton as long as only Kaluza--Klein type
couplings are considered.
The shaky point about this argument concerns the non--renormalizability
of the model under discussion and the question of the very meaning of
perturbation theory. One is on much safer ground if supersymmetry
can be employed to exclude a perturbatively generated dilaton potential \cite{DS},
but going below the SUSY scale we have left that safe harbor behind.
Nevertheless, it turns out that the reasoning is not in contradiction
with a 1--loop calculation as long as one relies on dimensional
regularization:

On the 1--loop level the effective potential is generated by 1--loop diagrams
with only axions and dilatons of vanishing momenta as external particles.
There appear three types of relevant tree level vertices in (\ref{treeL}):\\
--- For external axions with gauge bosons in the loop there is only one
relevant vertex:
\begin{equation}\label{v1}
i\delta_{jk}\frac{q^2}{8\pi^2 f_a}\epsilon_{\mu\nu\rho\sigma}p_1^\rho p_2^\sigma,
\end{equation}
where the gauge bosons at the vertex
have momenta $p_1$, $p_2$, polarizations $\mu$, $\nu$
and orientations $j$, $k$ in the Lie algebra. 
We have also taken out one factor of $i$ into the momentum conserving
 factor $(2\pi)^4i\delta(p_1+p_2+k)$, with $k$ denoting the 4--momentum
of the axion.\\
--- For external dilatons with gauge bosons in the loop
there are enumerably many vertices:
\begin{equation}\label{v2}
\delta_{jk}\frac{1}{(if_\phi)^n}(p_{1\nu}p_{2\mu}-\eta_{\mu\nu}p_1\cdot p_2).
\end{equation}
--- The corresponding axion--dilaton vertices are
\begin{equation}\label{v3}
-\Big(\frac{2i}{f_\phi}\Big)^n p_1\cdot p_2,
\end{equation}
where now $p_1$ and $p_2$ are the incoming axion momenta.

 From the vertices it is immediately clear that no perturbatively generated
axion potential appears in the theory: (\ref{v3}) vanishes due to the vanishing
axion momenta, while (\ref{v1}) vanishes due to momentum conservation
with vanishing external axion momentum.

On the other hand, a diagram with external zero--momentum dilatons
and axions or gluons in the loop is directly proportional to $\int d^4 p$,
and this vanishes in dimensional regularization \cite{mv}. After Wick rotation
one finds
\[
\int d^{4-2\varepsilon}p=\lim_{\alpha\to 0}
\int d^{4-2\varepsilon}p\,\frac{1}{(p^2+m^2)^\alpha}=
\lim_{\alpha\to 0}i\pi^{2-\varepsilon}m^{4-2\varepsilon-2\alpha}
\frac{\Gamma(\alpha-2-\varepsilon)}{\Gamma(\alpha)}=0
\]
and no $\phi^n$--vertices could be inferred from these diagrams.

\end{document}